\documentclass{sig-alternate-05-2015}

\usepackage{ifpdf}

\ifpdf
\else
\usepdflatexoryouhaveanerrornow
\fi

\usepackage[pass]{geometry}

\usepackage{endnotes}
\usepackage{times}
\usepackage[T1]{fontenc}
\usepackage{float}
\usepackage{graphicx}
\usepackage{array}
\usepackage{url}
\usepackage{comment}
\usepackage{framed}
\usepackage{multirow}

\usepackage{balance}

\usepackage{hyperref}

\usepackage{listings}

\usepackage{amsmath,amsthm,amssymb,latexsym, pifont}

\usepackage[usenames]{color}
\usepackage{ifsym}
\usepackage{wasysym}
\usepackage{booktabs}
\usepackage{mdwlist}
\usepackage{caption}
\usepackage{subcaption}
\usepackage{latexsym}
\usepackage{epstopdf}
\usepackage{color}
\usepackage{enumitem}

\usepackage{microtype}

\usepackage[ruled]{algorithm2e}

\usepackage{cite}
\usepackage{xspace}

\usepackage{xparse}
\usepackage{nccmath}

\newcommand{\xref}[1]{Section~\ref{#1}}
\newcommand{\cref}[1]{Chapter~\ref{#1}}

\newcommand{\fref}[1]{Fig.~\ref{#1}}
\newcommand{\tref}[1]{Table~\ref{#1}}

\newcommand{\first}{\emph{(i)}~}
\newcommand{\second}{\emph{(ii)}~}
\newcommand{\third}{\emph{(iii)}~}
\newcommand{\fourth}{\emph{(iv)}~}
\newcommand{\fifth}{\emph{(v)}~}
\newcommand{\sixth}{\emph{(vi)}~}
\newcommand{\seventh}{\emph{(vii)}~}
\newcommand{\eighth}{\emph{(viii)}~}

\newcommand{\ie}{i.e., \@}
\newcommand{\eg}{e.g., \@}

\newcommand{\cf}{cf. \@}

\newcommand{\etal}{et~al.\xspace}

\newcommand{\perc}{\,\%\xspace}

\definecolor{darkgreen}{rgb}{0,0.5,0}
\definecolor{brown}{rgb}{0.7,0.3,0}
\definecolor{darkblue}{rgb}{0,0,0.5}

\newcounter{fn1}
\newcounter{fn2}
\newcounter{fn3}
\newcounter{fn4}
\newcounter{fn5}
\hypersetup{pdfborder=0 0 0,
  colorlinks=true,
  citecolor=black,
  linkcolor=black,
  urlcolor=darkblue
}

\theoremstyle{definition}

\usepackage{authblk}

\let\underscore\_
\newcommand{\myunderscore}{\renewcommand{\_}{\underscore\hspace{0pt}}}
\myunderscore

\newcommand{\hide}[1]{}

\newcommand{\rscenario}[1]{\textsc{#1}\xspace}
\newcommand{\mountain}{\rscenario{Pointy Peak}}
\newcommand{\rollercoaster}{\rscenario{Unrestricted}}
\newcommand{\plateau}{\rscenario{Wide Peak}}
\newcommand{\podium}{\rscenario{With Steps}}

\newcommand{\VG}{V_{G}}
\newcommand{\EG}{E_{G}}
\newcommand{\bE}[1][e]{\ensuremath{\mathsf{bw}_{#1}}}

\newcommand{\lE}[1][e]{\ensuremath{\mathsf{lat}_{#1}}}

\newcommand{\Requests}{\mathcal{R}}

\newcommand{\sR}[1][R]{\ensuremath{\mathsf{s}_{#1}}}
\newcommand{\tR}[1][R]{\ensuremath{\mathsf{t}_{#1}}}
\newcommand{\bR}[1][R]{\ensuremath{\mathsf{bw}_{#1}}}
\newcommand{\lR}[1][R]{\ensuremath{\mathsf{lat}_{#1}}}
\newcommand{\lP}{\ensuremath{lat_{P}}}

\def\algMCF/{\ensuremath{\texttt{MinCostFlow}}}
\def\algMCA/{\ensuremath{\texttt{MinCostAssignment}}}
\def\algSP/{\ensuremath{\texttt{ShortestPath}}}
\def\algMSP/{\ensuremath{\texttt{MinAllShortestPath}}}

\def\algLP/{\ensuremath{\texttt{solveLP}}}
\def\algSolveSepSolve/{\ensuremath{\texttt{solveSeparateSolve}}}
\def\algAddLocalConstraints/{\ensuremath{\texttt{addConstraintsLocally}}}

\def\infeasibleLP/{\ensuremath{\texttt{infeasibleLP}}}
\def\objectiveLimit/{\ensuremath{\texttt{objectiveLimit}}}
\def\disableGlobalCutoff/{\ensuremath{\texttt{disableGlobalPrimalBound}}}

\def\NULL/{\ensuremath{\texttt{null}}}

\makeatletter
\def\@copyrightspace{\relax}
\makeatother

\begin{document}

\title{Stitching Inter-Domain Paths over IXPs}

\author{Vasileios Kotronis$^1$ \quad
Rowan Kl\"oti$^1$ \quad
Matthias Rost$^2$ \quad
Panagiotis Georgopoulos$^1$\vspace{-5mm}\\
Bernhard Ager$^1$ \quad
Stefan Schmid$^{3}$ \quad
Xenofontas Dimitropoulos$^{4,1}$\\
\small{$^1$ETH Zurich, Switzerland \quad\quad
$^2$TU Berlin, Germany \\
$^3$Aalborg University, Denmark \quad\quad
$^4$Foundation of Research and Technology Hellas (FORTH), Greece}\\
}

\maketitle

\begin{abstract}

Modern Internet applications, from HD video-conferencing to health monitoring and 
remote control of power-plants, pose stringent demands on network latency, bandwidth and
availability. Centralized inter-domain routing brokers is
an approach to support such applications and provide inter-domain guarantees,
enabling new avenues for innovation. These entities centralize routing control
for mission-critical traffic across domains, working in parallel to BGP. 
In this work, we propose using IXPs as natural points for stitching
inter-domain paths under the control of inter-domain routing brokers. 
To evaluate the potential of this approach, we first map the global
substrate of inter-IXP pathlets that
IXP members could offer, based on measurements for 229 IXPs worldwide.
We show that using IXPs as stitching points has two useful properties. 
Up to 91\% of the total IPv4 address space can be served by such
inter-domain routing brokers when working in concert with just a handful
of large IXPs and their associated ISP members. Second, path diversity on the inter-IXP graph increases
by up to \emph{29} times, as compared to current BGP valley-free
routing.
To exploit the rich path diversity, we
introduce algorithms that inter-domain routing brokers can use to
embed paths, subject to bandwidth and latency constraints.
We show that our algorithms scale to the sizes of the measured
graphs and can serve diverse simulated path request mixes.
Our work highlights a novel direction for SDN innovation across domains,
based on logically centralized control and programmable IXP fabrics.

\end{abstract}

\section{Introduction}
\label{sec:intro}

A great success of the Internet is that it has been used in ways that
were never anticipated during its early days. Carrying voice data\footnote{Increasingly, traditional 
telcos like Deutsche Telekom are planning to switch to IP telephony exclusively~\cite{dtswitchtel}.} and connecting
stock exchange markets are just two examples of such use cases.
Nothing suggests that this innovation will not persist in the future.
We see though that modern applications have increasingly tighter requirements for bandwidth, latency
and/or availability~\cite{bwtrends}. For example, real-time HD video streaming, telemusic~\cite{telemusic-LOLA},
remote control of critical infrastructure, such as power plants~\cite{power-outage-italy-article},
or even telesurgery~\cite{telesurgery} are emerging or envisioned
applications with strict network requirements. Presently, ISPs are able to provide
certain QoS guarantees~\cite{SprintGMLPSQoS} only in intra-domain settings based on technologies such as leased circuits and VPN
tunnels, \eg over MPLS-TE.~However, despite several
research and standardization efforts, providing QoS guarantees at the inter-domain level has seen very limited
success so far~\cite{QoSFail,QoSChallenges,4050103,IntScaleQoS}. Besides, current BGP routing can lead
to inefficient paths across domains, triangle inequality violations, and long-lasting outages~\cite{lumezanu2009triangle,
katz2008studying,HotInterconnects}.

During the last decade, an increasing number of proposals coming from diverse angles advocate
inter-domain routing brokers~\cite{RaaS,RouteSource,MINT,RouteBazaar,Zhang:2000:DQC:347057.347403,QoSIPMPLS}
as an approach to enable ISPs to cooperate and provide end-to-end (e2e) guarantees.
In these schemes, ISPs provide QoS-enabled pathlets~\cite{Pathlet}, which are stitched together by an
inter-domain routing mediator, \eg a bandwidth broker~\cite{MINT}.
Related initiatives are currently explored in the industry{~\cite{GEANT-BoD}
and in standardization bodies, in particular in the context of the PCE (Path Computation Element) architecture~\cite{PCEARch,PCEP,pace}.

This work visits logically centralized inter-domain mediators
in light of the evolving Internet ecosystem. Namely, the
Internet is becoming denser and more flat~\cite{FlatInternet,InterDomTraf,IXPStructure}
because public Internet eXchange Points (IXPs) are continuously rising in
number and size~\cite{ixp-anatomy,IXPsEye}.
In parallel, the paradigm shift towards network virtualization~\cite{flowvisor2}
and Software-Defined Networking (SDN)~\cite{openflow_paper}
introduces new possibilities in network management and innovation,
also in the context of IXPs, \eg as shown in the Software-Defined
eXchange (SDX) approach~\cite{SDX-SIGCOMM}. While SDX enables new
services at individual IXPs,  we focus on multi-IXP services.

\textbf{Contribution 1: Stitching inter-domain paths via IXPs.}
We propose using IXPs for stitching paths under the control 
of inter-domain routing brokers. We call such brokers 
\emph{Control eXchange Points} (CXPs)\footnote{CXPs can generally use any switching point between 
ISPs.}. The choice of IXPs
as switching points exploits their rich connectivity, enabling high path diversity and
global client reach with deployment in only a few well-connected IXPs.  
CXPs enable the utilization of additional path diversity compared to
current BGP-based inter-domain paths, which typically follow valley-free routing
policies~\cite{NoGlobalCoor,gill2013survey, giotsas2014complex-relationships}. 
CXP-stitched paths can freely cross multiple IXPs, yielding new paths that BGP hides.

\textbf{Contribution 2: Mapping the IXP Multigraph.} To evaluate the potential
of CXPs, we map the global Internet substrate for pathlet stitching over IXPs. 
In particular, we outline a novel abstraction of the Internet topology, in which
vertices are IXPs and edges are virtual links connecting two IXPs over an ISP. We call this abstraction
the \emph{IXP multigraph} because two IXPs can be generally connected with multiple edges over different ISPs.
This abstraction hides the internal details of an ISP (including the technologies
that can be leveraged to provide intra-domain QoS~\cite{Zhang:2000:DQC:347057.347403}),
and serves a clean separation of concerns between intra-
and inter-domain QoS routing that is consistent with the status quo. 
We analyze the member ISPs of 229 IXPs using data from Euro-IX~\cite{EuroIX}
and show that CXPs can service, \eg 40\perc of the globally announced IPv4 addresses
through only the 5 largest IXPs. This increases to 91\perc if we also consider the 1-hop
customers of the IXP members. Second, we show that by relaxing valley-free constraints,
CXPs can greatly increase path diversity by up to 29 times compared to BGP
valley-free routing.

\textbf{Contribution 3: Algorithms.}
We present algorithms to efficiently exploit the high path diversity
observed in the IXP multigraph. In particular, our algorithms aim at
maximizing the number of concurrently embedded paths, subject to
bandwidth and latency constraints. We describe online as well as hybrid
online-offline algorithms which sample feasible paths efficiently (i.e., in
polynomial time). These algorithms achieve different trade-offs between optimal
acceptance ratios and fast online computation, with the hybrid approach
realizing a balance between the two goals by reallocating paths in the
background based on an optimal offline algorithm. Using
simulation, we show that our algorithms scale to the sizes of the
measured graphs and derive insights on which variants should be
leveraged to serve diverse requests.

CXPs provide a possible avenue for SDN innovation at the inter-domain level.
In this context, we investigate both the algorithms that can serve as the controller logic of logically centralized inter-domain
routing brokers, operating on IXP multigraphs, and the interesting properties of this particular data plane. The latter
is studied both in space (incremental deployment at IXPs) and time, as the peering ecosystem evolves over the years. 
Moreover, we discuss further challenges for future work under the prism of a possible use case.

The rest of the paper is structured as follows. 
\xref{sec:use-case} provides the background on inter-domain service brokers and the motivation behind
our IXP-based approach.
\xref{sec:ixp-analysis} maps the global inter-IXP multigraph, based on Euro-IX and PeeringDB data, and characterizes its
high path diversity and client reach for inter-domain QoS.
\xref{sec:algo} presents algorithms for embedding paths
in IXP multigraphs and \xref{sec:eval} evaluates these
algorithms based on a custom simulator.
\xref{sec:disc} discusses our work under the prism of telesurgery as a use case, while 
\xref{sec:related} presents related literature.

\section{Service Brokers, IXPs and CXPs}
\label{sec:use-case}

This section first gives an overview of previous research on
centralized path brokers for inter-domain guaranteed services.
Second, we discuss why IXPs are suitable locations for
deploying the data plane elements of path brokers. Lastly, we
describe in detail the properties of our IXP-based path brokers,
which we call Control Exchange Points (CXP).

\subsection{Network Service Brokers}
\label{sec:brokers}

Previous research has focused on
bandwidth brokers for mediating the
concatenation of multiple guaranteed bandwidth 
pathlets (\eg MINT~\cite{MINT}), or for scaling up the
support for guaranteed bandwidth services within an ISP network (\eg the work of
Zhang \etal~\cite{Zhang:2000:DQC:347057.347403}).
Similar initiatives have created
bandwidth markets and commercial brokers, such as Geant's multi-domain Bandwidth-on-Demand service~\cite{GEANT-BoD}.
Other proposals introduce ``route bazaars'' between ISPs and end-users~\cite{RouteBazaar},
where pricing mechanisms and interactions directly affect path establishment.
Routing-as-a-Service controllers~\cite{RaaS}
have been proposed as potential broker implementations. 
Others have proposed entirely outsourcing routing control to
inter-domain SDN controllers~\cite{RouteSource}.  Such controllers
can deal with end-to-end path stitching using their bird's eye
view over the participating domains;
dynamic traffic management applications can operate on this global view.
Centralized routing controller platforms based on the Path Computation
Element (PCE) architecture~\cite{PCEARch, PCEP}
have been evaluated in the context of
QoS routing schemes for high capacity optical networks~\cite{PCEQoSPerf}.
The initial multi-domain intention of PCE was to help coordinate path
establishment requests, and to be able to compute an end-to-end path using
cooperative per-domain PCEs.
Systems like PCE are highly relevant for the implementation of brokers and
routing controllers, \eg applied on IP/MPLS domains ~\cite{QoSIPMPLS}, and are backed
up by IETF standardization efforts~\cite{PCEARch, PCEP}.

\begin{figure}[t]
\begin{center}
  \includegraphics[width=1.0\columnwidth]{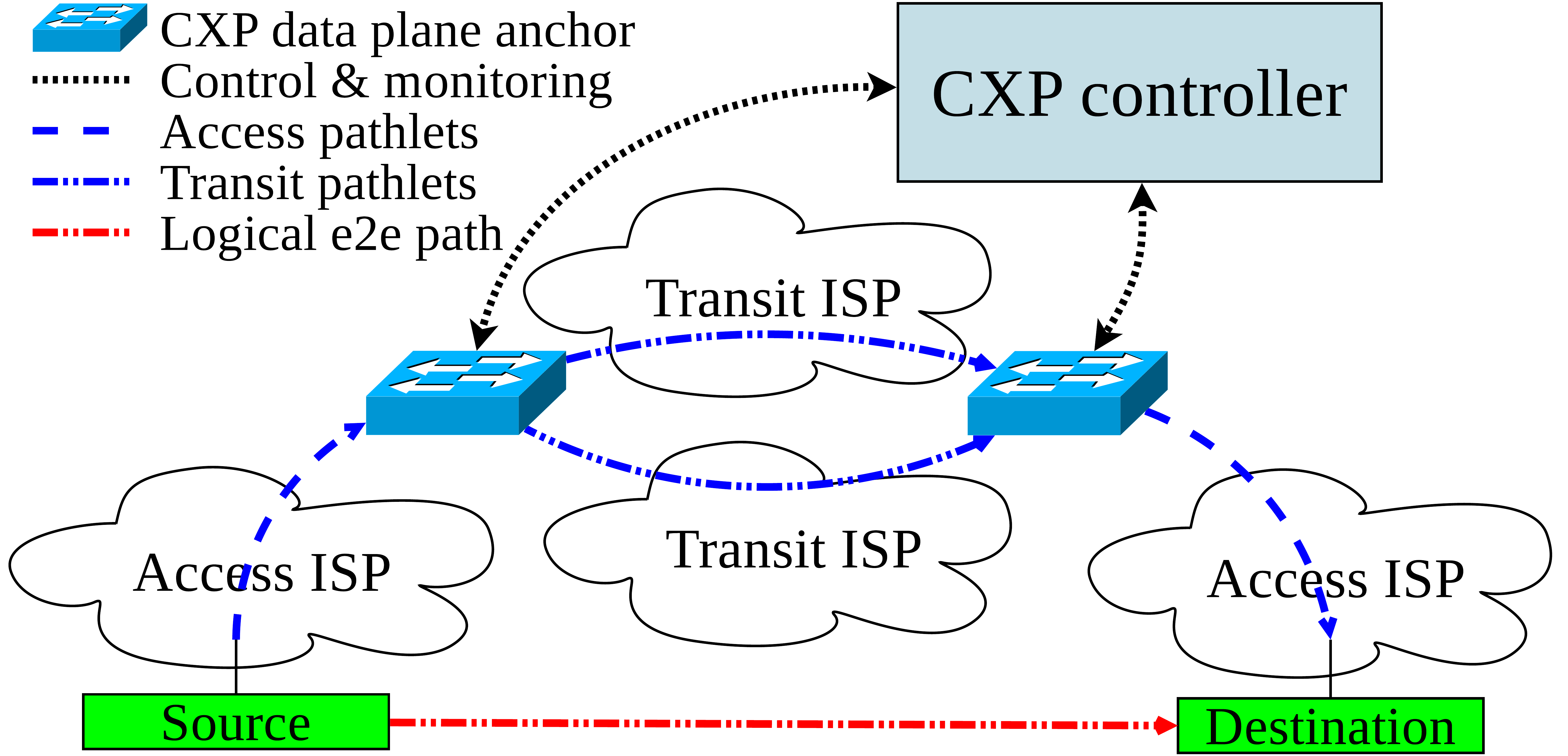}
	  \caption{The CXP stitches QoS-enabled e2e paths.}
	\label{fig:cxp-concept}
    \vspace{-20pt}
\end{center}
\end{figure}

\subsection{Deploying Service Brokers on IXPs}
\label{sec:IXPdep}

Brokers and controllers for guaranteed e2e services
need to exert inter-domain control through programmable
data plane elements, such as OpenFlow switches.
We call these elements \emph{anchors}, since they ``anchor''
inter-domain traffic switching to specific locations, decoupled
from the traffic management within \eg ISP domains.
The ideal anchor is adjacent to multiple
geo-diverse ISPs, is provisioned for high bandwidth and
availability, and is independent from a single ISP.
We observe that IXPs have all these properties and thus
provide ideal starting points for deployment.
IXPs are presently the hubs of multiple services surpassing their initial
goal of pure layer-2 switching fabrics~\cite{IXPsEye}: \first hosting route servers for ease of
BGP-based peering~\cite{richter2014peering}, \second mobile peering with 3G
providers for traffic convergence~\cite{AMSIX-GRX}, or \third the adoption of
SDN approaches for new inter-domain applications~\cite{SDX-SIGCOMM}---such as application-specific peering---are just a few examples.
They are therefore open to hosting new services for their members, together with increasing their
peering base. 

Modeling IXPs as vertices and inter-IXP pathlets as edges, the
resulting topology is a dense \emph{multigraph}: two IXPs can be
connected via multiple ISPs. This is quite common because many ISPs
are present at multiple different IXPs in parallel (\cf
\xref{sec:ixp-analysis} for details).
We base our study on this simple yet powerful observation,
enabling us to build a novel IXP-centric abstraction of the Internet
topology. Endpoints can connect to this topology via pathlets offered by
their access providers towards adjacent IXPs (see \fref{fig:cxp-concept}).

\subsection{CXPs}

Following the observation that IXPs provide ideal locations for data
plane anchors, we introduce \emph{Control Exchange
  Points} (CXPs), \ie control points which stitch pathlets
across multiple administrative domains to construct global paths. 
Here we discuss in detail how CXPs would operate and the existing or emerging
control and data plane technologies a CXP implementation could rely on. 
We note that the full implementation of a CXP is beyond the scope of this work.

\textbf{Basics.} A CXP is a logically centralized entity, applying inter-domain control
over how parts of Internet traffic are routed. In this context, it can, for
example, provide e2e QoS or support multicast services by selecting (a multitude of)
appropriate paths. A CXP works in parallel to
traditional routing and can control parts of
traffic independently from BGP, \eg utilizing flow space isolation
mechanisms~\cite{flowvisor2}. CXPs use
data plane anchors which classify and switch traffic, such as SDN
switches~\cite{openflow_paper}. 
Software Defined Internet eXchanges (SDX)
as proposed by Gupta \etal~\cite{SDX-SIGCOMM} could constitute an IXP-based deployment
possibility. CXP control planes can be built using
PCEs~\cite{PCEARch}. PCEs can reduce the required inter-domain signaling, enforce traffic
access policies and hierarchically manage multi-technology domains.
Moreover, a potential cooperation between IXP Route Servers
and PCEs could enable CXPs to respond dynamically
to changing requirements over a set of IXP-mediated inter-domain connections.
Besides public IXPs, anchors can be deployed at private peering
points for augmenting geographical coverage, if required.
Between data plane anchors, traffic is
shipped on virtual links which are parts of e2e paths and act as \emph{pathlets}~\cite{Pathlet}.
Pathlets are provided by ISPs and may be annotated with specific properties,
such as bandwidth and latency guarantees (if QoS is to be supported), with simple connectivity
as the baseline. When a client requests an e2e path, the CXP
has to find a suitable sequence of pathlets that meet the client's QoS
requirements.

\textbf{Providing Pathlets.} Pathlets can be provided by ISPs with existing tunneling
techniques, such as MPLS, GRE and VPNs, or emerging SDN
approaches based on flow space allocation along a network path~\cite{flowvisor2, openflow_paper}.
Within the ISP backbone, QoS guarantees are provided via traffic engineering and
prioritization techniques~\cite{TEPrinciples, Zhang:2000:DQC:347057.347403}. MPLS-TE~\cite{MPLSQoS} is one example technology.
The ISP is responsible for providing
cross-traffic isolation internally, keeping its management policies
confidential. The CXP on the other hand, provides isolation on
the data plane anchors.
An ISP may provide multiple pathlets between two data plane
anchors with different properties for service differentiation
or fail-over. We note that CXPs do not have control over how \emph{physical
pathlet redundancy} is achieved within the ISP. Availability properties
(\eg for telesurgical applications) should therefore accompany the ISP-originated
pathlet advertisements. One way to achieve this is by annotating pathlets
with Shared Risk Link Group (SRLG) IDs~\cite{SRLG}.
The incentive for ISPs to provide pathlets is the revenue generated
when their pathlets are used for e2e services; any ISP can be a provider. As shown in \fref{fig:cxp-concept}, the ISPs of
the source and the destination offer access pathlets to connect to
ISP-adjacent data plane anchors, while the intermediary ISPs offer transit pathlets over their
domains, between anchors.

\textbf{CXP Tasks.} The CXP \first handles new requests for QoS-enabled
paths (admission control), \second computes and sets up suitable paths (embeddings), \third monitors pathlet
availability and compliance with QoS guarantees, and \fourth performs
reembedding, if required. A client negotiates her request directly
with her access ISP, which selects a suitable CXP for establishing the
inter-domain route out of a set of available CXPs. The ISP forwards the client's request to the chosen CXP
which in turn computes a suitable e2e path. The CXP reserves
capacity on the selected pathlets and then configures the respective
data plane anchors. Accordingly, the client's ISP has to configure its
network such that the quality sensitive traffic is sent via a pathlet to the correct
data plane anchor. A CXP monitors the bandwidth, latency and availability of a path for
the duration of the client's reservation, using existing technologies and approaches~\cite{bwest, InternetDelay, QoSMon}.
If the client's requirements are violated or a
pathlet becomes unavailable, the CXP chooses and configures an
alternative path for the affected part(s) of the traffic; this can even be
a ``hot-standby'' backup path carrying traffic duplicates.
Besides, the CXP may choose to better
utilize the available pathlets by re-embedding paths and defragmenting
the substrate resources.

\newpage
\section{The IXP Multigraph}
\label{sec:ixp-analysis}

In this section we measure and characterize the inter-IXP multigraph,
\ie the substrate on which inter-domain path brokers may operate. 
This analysis is necessary to understand
where inter-domain control could be applied as well as the efficiency of
incremental deployment, and is complementary to research related to scaling up
CXP-like control planes~\cite{onos} or investigating the trade-offs involved in logical
centralization~\cite{LogCen}.
We thus answer the following questions:
\first how many IXPs need to participate so that CXPs can provide guaranteed services to a large
population of the Internet, assuming that their member ISPs would offer the necessary pathlets,
and \second how much path choice and diversity we can gain compared to classic BGP routing practices.
We highlight this because currently, due to valley-free routing~\cite{NoGlobalCoor} and the prevalence of
peer-to-peer links at IXPs~\cite{ixp-anatomy}, Internet paths
normally cross at most one IXP. IXP-based path brokers simplify the use of paths that cross 
multiple IXPs.

\begin{table}[t]
\centering
\small
\tabcolsep4.5pt
\begin{tabular}{l*{6}r}
\toprule
 & \multicolumn{6}{c}{Scale-Down Factor (SDF)} \\
\cmidrule{2-7}
Property                   & 1       & 2       & 4        & 8       & 16         & 32 \\
\midrule
Node count                 & 229     & 115     & 57       & 28      & 14         & 7 \\
Edge count                 & 49k     & 29k     & 15k      & 6.5k    &  3.9k      & 1.1k \\
Diameter                   & 5       & 5       & 3        & 2       & 2          & 1 \\
Av. node degree            & 220     & 250     & 260      & 230     & 280        & 160 \\
Av. edge multiplicity      & 4.3     & 6.0     & 8.3      & 12.     & 25.        & 26. \\
Av. shortest path len.     & 1.9     & 1.6     & 1.4      & 1.3     & 1.1        & 1.0 \\
Av. clustering coeff.      & 0.80    & 0.82    & 0.85     & 0.87    & 0.93       & 1.0 \\

\bottomrule
\end{tabular}
\caption{Properties of the graphs generated from the Euro-IX dataset at
various scale-down factors (SDF); larger SDFs correspond to smaller CXP penetration and vice versa.}
\label{tab:graph-properties}
\end{table}

\subsection{Mapping the Inter-IXP Topology}

We use four datasets to map the inter-IXP topology and the IPv4 address space: \first the Euro-IX~\cite{EuroIX}
and \second PeeringDB~\cite{PeeringDB} databases, from which we obtained IXP membership data,
\third the CAIDA AS relationship
data~\cite{caida-as-rel, luckie-as-relationships-2013}, and \fourth the CAIDA RouteViews
AS-to-prefix data~\cite{caida-routeviews}. Due to space constraints we
report results only for Euro-IX, which also provides geographic coordinates
of IXPs (used to determine distances between IXP locations in \xref{sec:eval}) in contrast
to PeeringDB. Analysis on PeeringDB data
further corroborates our findings. We note that, in general, there are multiple publicly available
sources of information on IXPs, including Euro-IX, PeeringDB, PCH, IXP websites and public data
from BGP route collectors. For a comprehensive comparison of these sources in terms of completeness and
accuracy we refer the reader to the work of Kl\"oti \etal~\cite{kloti2016ixps}, which serves as complementary research to
the investigation of the properties of such datasets. 

Using a snapshot of the Euro-IX peering database~\cite{EuroIX}, we
extracted membership data for 6,542 ASes in 277 IXPs. After ignoring IXPs which
had no members or had only members which advertised no IP prefixes, we have 6,122 ASes in 231 IXPs. Two further IXPs which have no
connections to others are discarded. The final (connected) graph consists of 229 IXPs
and
$\sim$49k edges between IXPs, crossing ISPs that peer concurrently with these IXPs.
We derive simple graphs by collapsing multi-edges to single edges, annotated with the
initial edge multiplicity.

We scale down the extracted inter-IXP topology assuming that a CXP does not have all the IXPs
at its disposal, but gradually recruits IXPs to maximize the IP address space it can serve.
Each new IXP provides access to more client address space served by its member ISPs.
We determine a suitable order based on a greedy heuristic, starting with the IXP having the largest
address space coverage and in each iteration adding the IXP which
yields the greatest number of non-overlapping addresses. We assume that whenever
we add a new IXP, all its member ISPs would host pathlets that: \first connect their edge
clients to the new IXP (via \emph{access} pathlets, cf. \fref{fig:cxp-concept}) , and \second connect
the new IXP to other CXP-enabled IXPs at which these ISPs are present
(via \emph{transit} pathlets, cf. \fref{fig:cxp-concept}).
We make this assumption, since our goal is to investigate the potential of an IXP-centric multigraph for CXP deployment,
as the CXP approaches more and more IXPs.
Each IXP is associated with an ISP membership base, which we want to examine in full. 
The dynamics of the pathlet market will eventually determine
which IXPs and ISPs will participate, which pathlets they will advertise and which clients
will choose to connect under diverse QoS guarantees. For such market analyses,
investigating pathlet pricing and ISP participation,
we refer the reader to
works such as MINT~\cite{MINT} or RouteBazaar~\cite{RouteBazaar}.

\subsection{Properties of the Inter-IXP Multigraph}
\label{sec:anatomy}

\tref{tab:graph-properties} gives an overview of the properties of the
inter-IXP multigraph at different scales. The scale-down factor 32 corresponds to a
small CXP deployment on 7 IXPs, while a factor of 1 involves all the 229
IXPs. We first observe average shortest path lengths between 1 and
1.9 edges. This observation combined with the high clustering factors suggests small world properties.
Furthermore, multi-edges result in very high average node degrees, \eg of
160 in the initial topology with 7 IXPs.
\fref{fig:common-as} shows the Complementary Cumulative
Distribution Function (CCDF) of the edge multiplicity, \ie the
number of parallel ASes that connect pairs of IXPs, in the full (unmodified)
topology. We observe that a few pairs of IXPs are
interconnected by over a hundred distinct ASes, each of which is in a
position to offer one or more pathlets between each pair.
Between the largest IXPs, which form the most likely
targets for an initial deployment, hundreds of pathlets---over member ISPs---may be available.

\fref{fig:path-div} shows the CCDF of \emph{path diversity}, which is the number
of \emph{edge-disjoint} paths between each pair of IXPs, computed with the minimum cut.
These paths can cross multiple IXPs and may be composed of multiple pathlets used in sequence.
Conceptually, the cut provides the minimum number of pathlets which would have to
be removed so that no path \emph{at all} is found between these IXPs.
We note however, that a failure inside a single ISP (\eg related to internal routing)
can affect many pathlets offered by this ISP. Also, different ISPs may share the same physical cables (\eg transatlantic fiber links). 
As \fref{fig:common-as} and \fref{fig:path-div} show,
the path diversity is much higher than the direct connectivity \ie edge multiplicity between pairs of IXPs.
Thus even when \emph{all} direct ISP pathlets between an IXP pair fails, multiple indirect paths crossing other ISPs
and IXP anchors may be used to replace the lost connectivity.

\begin{figure}[t]
\centering
  \begin{subfigure}[b]{0.49\columnwidth}
  \includegraphics[trim= 10mm 10mm 0mm 12mm,clip=true,width=1.0\columnwidth]{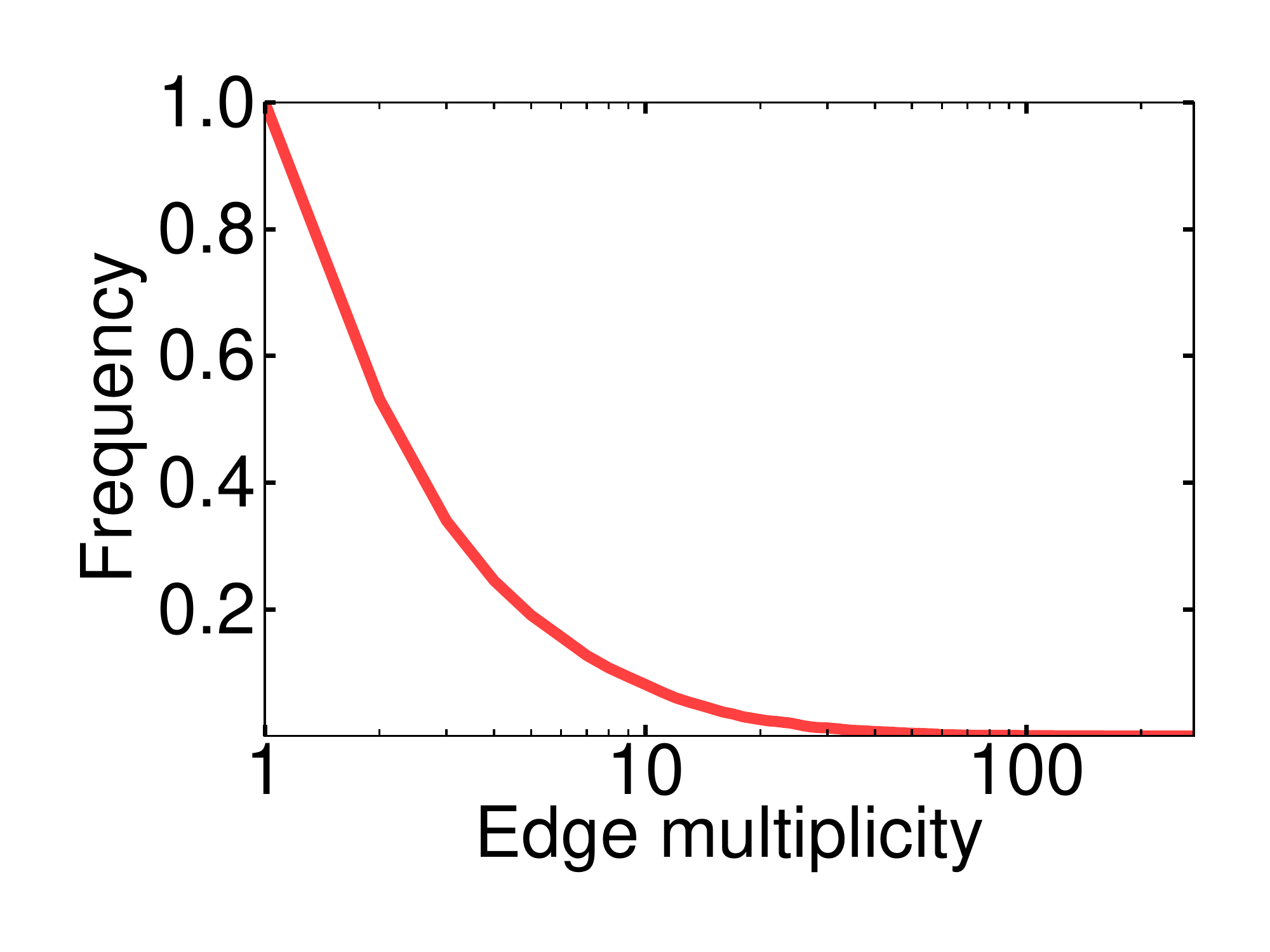}
\caption{}
\label{fig:common-as}
  \end{subfigure}
  \begin{subfigure}[b]{0.49\columnwidth}
  \includegraphics[trim= 10mm 10mm 0mm 12mm,clip=true,width=1.0\columnwidth]{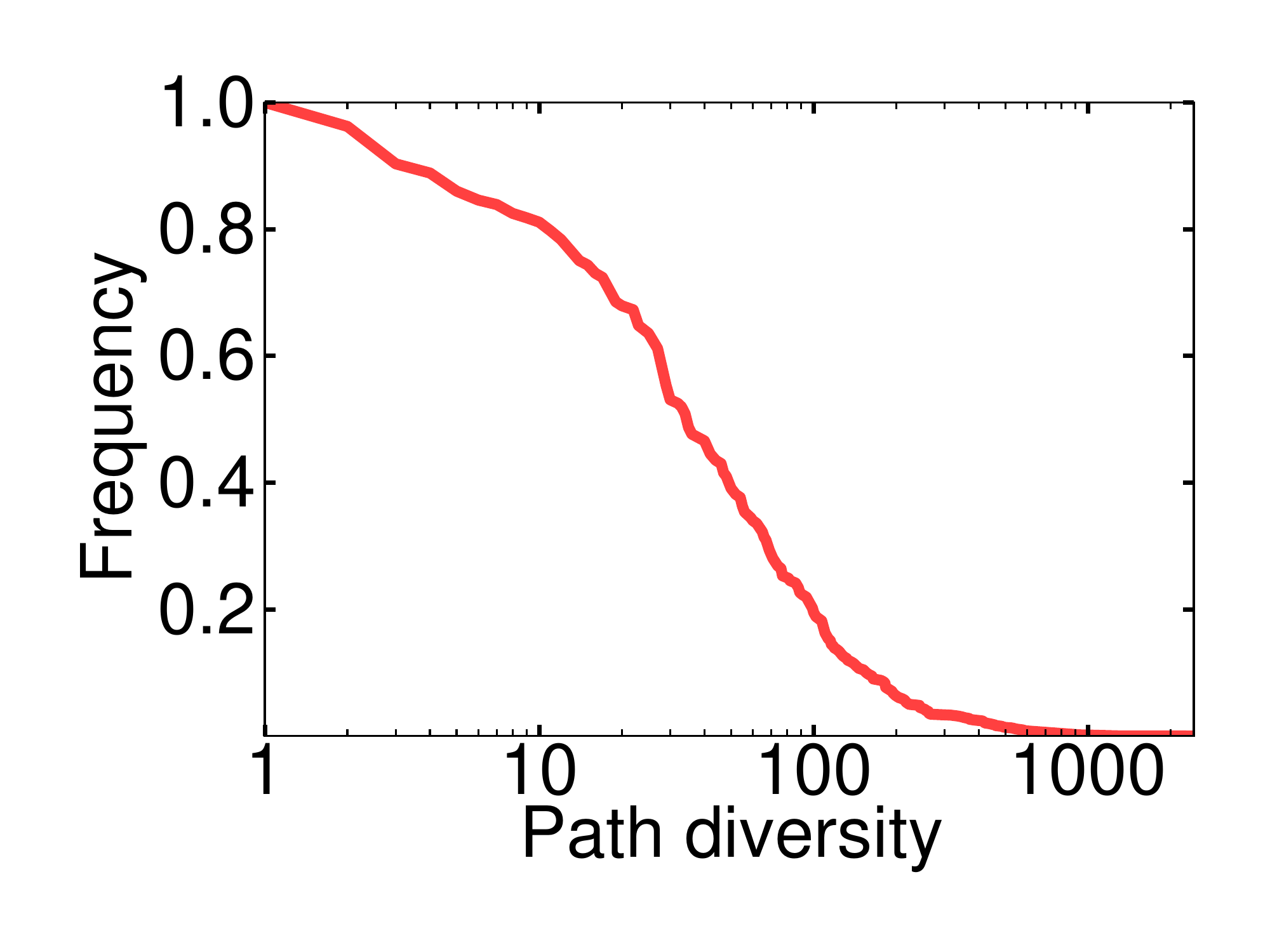}
\caption{}
\label{fig:path-div}
  \end{subfigure}
\caption{CCDFs of edge multiplicity and path diversity}
\end{figure}

\subsection{Reaching Clients with a Handful of IXPs}
\label{sec:scaling}

To be successful, reaching a large client base is important for a CXP.
Therefore, we address the question of how much of the IPv4 address
space can be reached from IXPs and their members.  \fref{fig:address-coverage} depicts the IP
address coverage versus the number of participating IXPs, assuming a
greedy strategy maximizing IP address coverage. We show results both
for directly adjacent IXP members as well as those connected over a single
intermediate ISP (one hop). We observe that we can serve over 1 billion IP
addresses via only 5 CXP anchors in well-connected IXPs for directly
connected customers, which is 40\perc of the announced IPv4 addresses in the Internet.
This increases to 2.4 billion IP addresses (91\perc of announced addresses) if we also
consider the 1-hop customer cone of the IXP members.
With 20 IXPs, more than 1.5 billion IP addresses (>50\perc of announced addresses) can be
reached directly. This allows an initial deployment of just a few
IXPs to serve large parts of the IPv4 address space and enables efficient incremental adoption
of inter-domain QoS-enabled services.
Further use of private peering points might selectively augment the required coverage, where applicable.

\begin{table}[t]
\centering
\scriptsize
\tabcolsep3.5pt
\begin{tabular}{ll*{6}r}
\toprule
               & & \multicolumn{6}{c}{Perc. of added p2p links}\\
\cmidrule(rl){3-8}
               & & \multicolumn{2}{c}{0\perc} & \multicolumn{2}{c}{25\perc} & \multicolumn{2}{c}{50\perc}\\
\cmidrule(rl){3-4}\cmidrule(rl){5-6}\cmidrule(rl){7-8}
Scenario       & Description              &  $\mu$ & $\mathrm{M}$ &   $\mu$ & $\mathrm{M}$ &  $\mu$ & $\mathrm{M}$ \\
\midrule
\mountain      & Valley-free              &    2.9 &            2 &     3.2 &            2 &    3.3 & 2 \\
\plateau       & + multiple peering links &   10.  &            2 &    43.  &            3 &   70.  & 3 \\
\podium        & + unconstrained peering  &   19.  &            3 &    68.  &            4 &   104.  & 4 \\
\rollercoaster & No restrictions          &   42.  &            5 &    108.  &            7 &  143.  & 7 \\
\bottomrule
\end{tabular}
\caption{AS-level policy models and their mean ($\mu$) and median ($\mathrm{M}$) path diversity, with
added p2p links.}
\label{tab:as-path-diversity}
\end{table}

\subsection{Rich Policy-Compliant Path Selection}
\label{sec:policies}

We next evaluate the increase in path diversity gained when using
a CXP-enabled IXP multigraph with relaxed peering policies as compared to valley-free
routing of the AS-level topology. 
The most constrained policy corresponds to the
traditional valley-free model~\cite{NoGlobalCoor} (scenario \mountain);
this allows the sequential composition of an uphill path (over customer-to-provider links), then at most one peer-to-peer (p2p)
link, and a downhill path (over provider-to-customer links), resembling a mountain with a rather narrow peak. The upper bound
on path diversity is achieved with the
unrestricted policy scenario (scenario \rollercoaster).
We investigate two additional scenarios by gradually relaxing the valley-free conditions.
\first The \plateau scenario extends valley-free routing by allowing
an arbitrary number of p2p hops between the uphill and the
downhill path, instead of at most one, representing a scenario where
there is exactly one CXP-mediated path traversed, passing over multiple IXPs.
\second
The \podium scenario allows an unlimited number of p2p links anywhere in the uphill path,
and also in the downhill path. Any number of CXP-mediated paths can be traversed either while
climbing uphill or descending downhill; this results
in a step-wise setup, \ie a mountain with potentially wide plateaus at different altitudes.

To address the known deficiency in detecting p2p links using the
current methodology to find AS-level links~\cite{ixp-anatomy}, and to investigate
the effect of more extensive peering on the Internet topology, we augment the AS
relationship graph with p2p links derived from IXP membership.
A given percentage of the derived links (\cf \tref{tab:as-path-diversity}) is added to the graph,
chosen uniformly at random; gradual addition is depicted with increasing
percentages\footnote{Larger percentages were not investigated due to
the memory limitations of the current NetworkX~\cite{NetworkX} min-cut implementations.}.
We estimate the corresponding policy-compliant AS-level path diversity,
capitalizing on our prior work~\cite{PolPath}. 
We use a sample size of 10K pairs of AS endpoints,
selected randomly, with each AS weighted by the number of IPv4 addresses it announces over BGP.

\tref{tab:as-path-diversity} shows the mean and median path
diversity observed for the various models and amounts of added p2p
links, while \fref{fig:path-div-AS} shows the distribution of path
diversity for the models without added p2p links.
We observe that transitioning from \mountain
to \plateau greatly increases the path diversity, even without
added p2p links. \plateau clearly has an advantage over
\mountain even when the latter has many new links added and the former
does not. This is true for the mean, but also the median, which
is less affected by the highly skewed distribution; for example, for tier-1 and
large tier-2 ISPs we see an increase by
up to a factor of 29. The \podium scenario has more modest gains
in median path diversity and lies within a factor of two of \rollercoaster ,
which is the upper bound. After examining the data, we observed that the advantage of \rollercoaster and \podium
over \plateau stems mainly from a relatively small number of very
well connected nodes. We therefore conclude that \first
relaxing constraints on peering policy greatly increases path
diversity, more so than simply introducing new p2p links, and \second
further relaxations of the model yield relatively modest benefits.
Lastly, the small world properties of the Internet AS-level topology graph, also
observed in the IXP multigraph abstraction (cf. \tref{tab:graph-properties}),
and our analysis of shortest path lengths show the following.
Since the Internet is densely
connected on the AS level, with the number of interconnections growing within a
valley-free regime, relaxing the
policy constraints does not yield \emph{shorter} paths
but simply allows us to use \emph{more} paths. We observed average lengths
within 3-4 hops irrespective of policy, in agreement with other related reports~\cite{ASlen}.

\begin{figure}[t]
\centering
\begin{subfigure}[t]{0.49\columnwidth}
  \includegraphics[trim= 10mm 5mm 5mm 20mm, width=1.0\columnwidth]{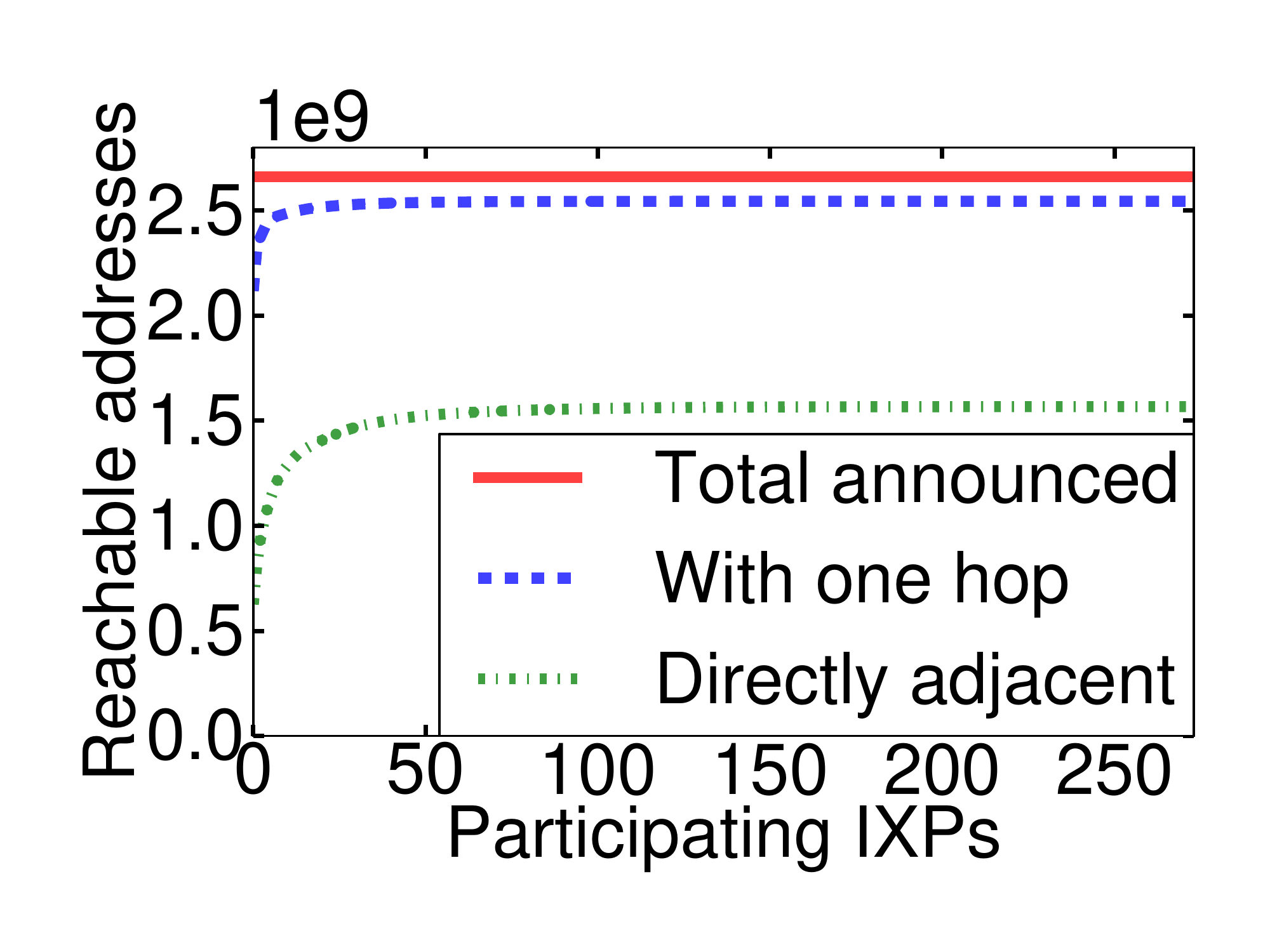}
\caption{\parbox{0.88\columnwidth}{Cumulative coverage of IPv4 address space originated by IXP members vs. participating IXPs}}
\label{fig:address-coverage}
\end{subfigure}
\begin{subfigure}[t]{0.49\columnwidth}
  \includegraphics[trim= 10mm 10mm 5mm 20mm, width=1.0\columnwidth]{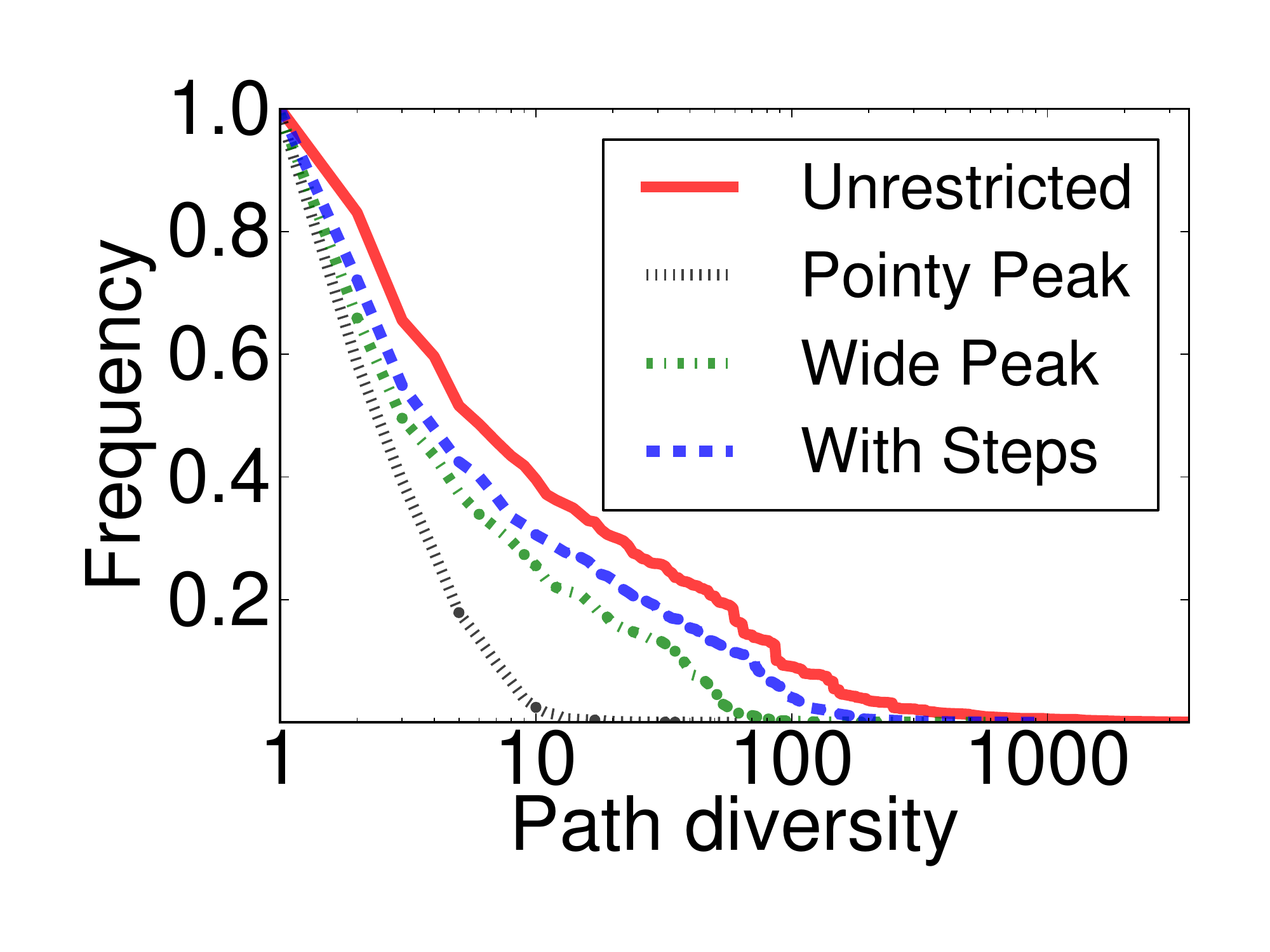}
\caption{\parbox{0.88\columnwidth}{CCDF of AS-level path diversity by scenario (with no added p2p links from IXP membership)}}
\label{fig:path-div-AS}
\end{subfigure}
\caption{CXP potential via IXP deployment}
\end{figure}

\subsection{Temporal Analysis of PeeringDB Graphs}
\label{sec:cxp-time}

In this section, we use available snapshots from the PeeringDB
database, complementary to the Euro-IX snapshot-based analysis,
in order to verify that our observations regarding the
properties of the projected CXP multigraph are valid \emph{over time}. 
We note that this analysis is not intended to be exhaustive,
but rather an indicative demonstration of the temporal evolution of the peering
ecosystem and the associated IXP multigraph, on which CXPs may operate. By knowing the
past, we can extrapolate what may happen in the future, as CXPs expand within an IXP-based Internet. 
For our temporal analysis we use monthly snapshots from crawling the PeeringDB website
over the months 3/2014 to 1/2015, effectively covering the monthly evolution of the data 
during the year 2014.
We also process the data extracted from SQL dumps on an almost yearly basis over 2008-2012.

We started with the evolution of the total number of the IXPs and ASNs which
participate in the peering ecosystem, over time. We observed that
the number of IXPs has been linearly increasing at a rate of $\sim$36 IXPs/year between the start of 2008 and
the end of 2013, while we witnessed an acceleration to a $\sim$115 IXPs/year rate of increase between the
start of 2014 and the end of 2014. The latter is a result of the recent influx
of small IXPs mostly located in South America, Africa and Australia; we will later revisit these IXPs
to determine their impact on the CXP multigraph. On the other hand, the number of ASNs that are reported
in PeeringDB seems to follow a steady linear increase at a rate of $\sim$460 ASNs/year.
Some of these ASes, as we show later, may be capable of acting as inter-IXP pathlet providers, thus contributing
to the density of the multigraph. In general, we observe that IXPs and their connected AS peers are rising
monotonically in sheer numbers over the years; IXPs have increased from less than 200 in the beginning of 2008
to more than 500 in the end of 2014, while the participating ASes have increased from $\sim$900 to $\sim$4000.

We next formed the actual corresponding IXP multigraph instances over time, and examined their sizes in terms
of nodes and edges. We observed that the number of IXP nodes in the multigraph is increasing at 
a rate of $\sim$32 IXPs/year. We note here that this behavior is a bit different than the one that we observed for 
\emph{all} the IXPs (nodes or not).
This is because the multigraph is based on the largest connected component of the IXP-based full graph; some of the IXP
nodes may be left out in case their member ASes cannot connect them to the rest of the multigraph.
Examples of such IXPs are the ones in some remote parts of Africa, Australia, East Asia and South America. 
Larger ISPs
that may peer concurrently at multiple IXPs around the world are usually not members of such small IXPs---at least
in the beginning.

Moreover, we observed that the number of
inter-IXP edges in the connected multigraph has been increasing at a rate of $\sim$4.8k edges per year between the years 2008 and the
third quarter of 2013, while afterwards the increase reaches a rate of $\sim$11.3k edges per year. 
By correlating this observation with the numbers of IXPs per ASN,
we deduce that the responsible ASes for this increase is the upper 1\% of all ASes. Each of these ASes is connected to at least 20 IXPs,
thus contributing at least 190 edges in the multigraph. The upper 0.1\% contributed at least 600 edges per ASN in 2008,
and at least 2.5k edges per ASN in 2014. This is probably due to their more aggressive peering 
at geo-diverse public IXPs in the recent years. In total, the number of edges has evolved from $\sim$10k
edges in 2008 to over 50k edges marking the start of 2015. Further correlation with the numbers of IXPs per ASN shows that
the multigraph has a ``slow'' changing component increasing at $\sim$5k edges per year; the lower 50\% of all ASes do not contribute at
all to this component, while the upper 50\% is responsible for sustaining this rate over the years. The upper 1\% of the highly connected ASes
is much more dynamic, contributing an extra $\sim$6k edges/year.

In \fref{fig:pdb-edge-mult-perc}
we examine the number of edges between \emph{directly connected} IXP pairs.
We observe that 50\% of the directly connected IXP pairs in the multigraph have an edge
multiplicity of 1, which is the typical median value. These pairs are connected via a single carrier ISP,
while each IXP of such edges can be connected to many other IXPs via different ISPs, albeit with a low redundancy. 
As we will show later, this behavior is balanced by the indirect path diversity and high redundancy in terms of indirect
paths between the IXP nodes.
In particular, as opposed to the low redundancy of these pairs, the remaining 49.5\% of the directly connected IXP pairs have a multiplicity ranging from
2 to 50. We note that the upper 0.5\% reaches levels of more than 50 edges per pair, with the
top 0.1\% striking an increasing multiplicity of over 100 in 2008, to over 300 in 2014.
By manual checking, we discovered that these pairs correspond to the largest global IXPs,
such as DE-CIX, AMSIX and LINX, connected over large shared ISP peering bases.

\begin{figure}[t]
  \centering
  \begin{subfigure}[b]{0.49\columnwidth}
    \includegraphics[trim= 0mm 0mm 0mm 0mm, width=1.0\textwidth]{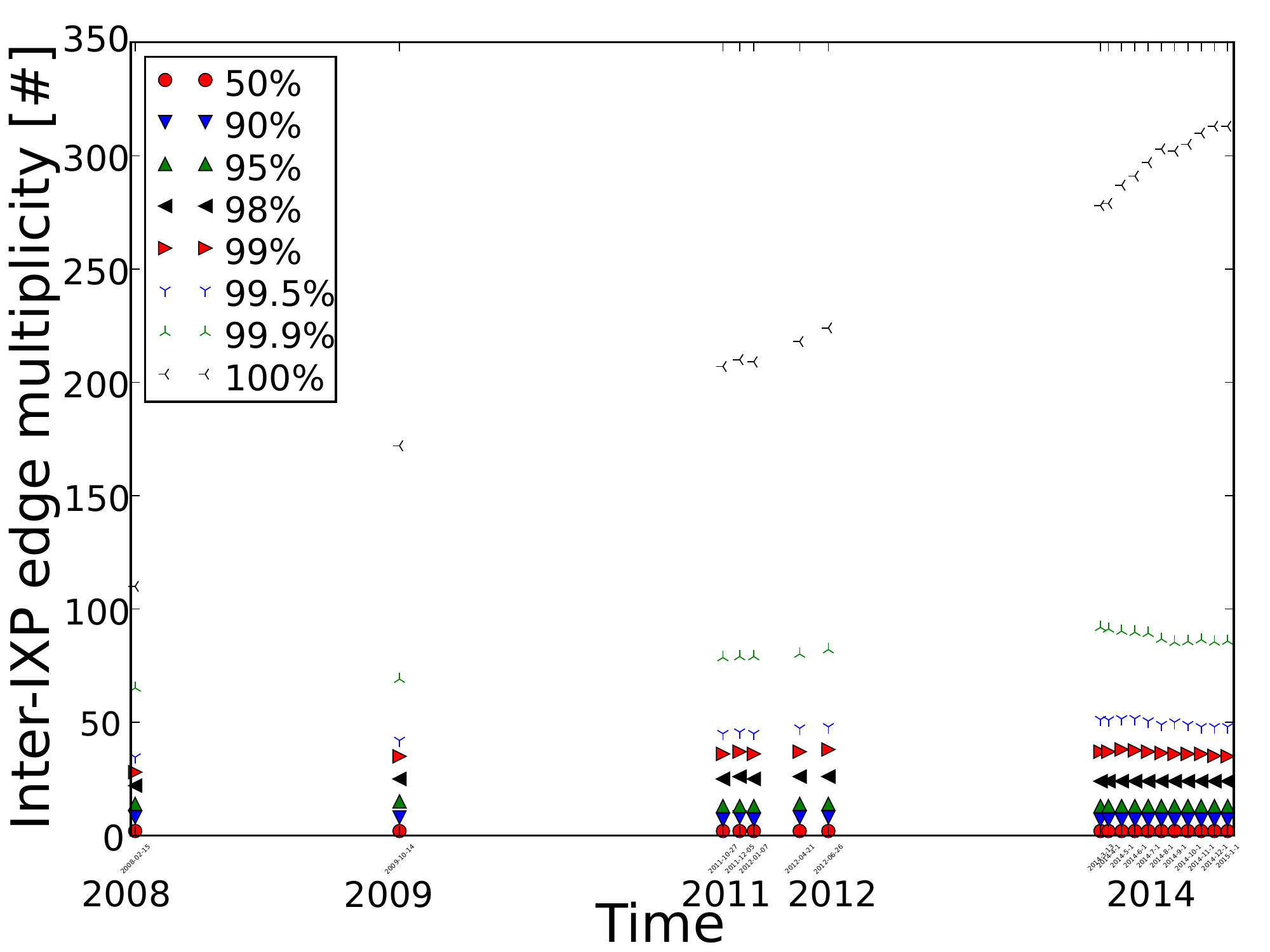}
    \caption{Edge multiplicity distribution percentiles between all the directly connected \emph{(IXP-IXP)} pairs.}
    \label{fig:pdb-edge-mult-perc}
  \end{subfigure}
  \hfill
  \begin{subfigure}[b]{0.49\columnwidth}
    \includegraphics[trim= 0mm 0mm 0mm 0mm, width=1.0\textwidth]{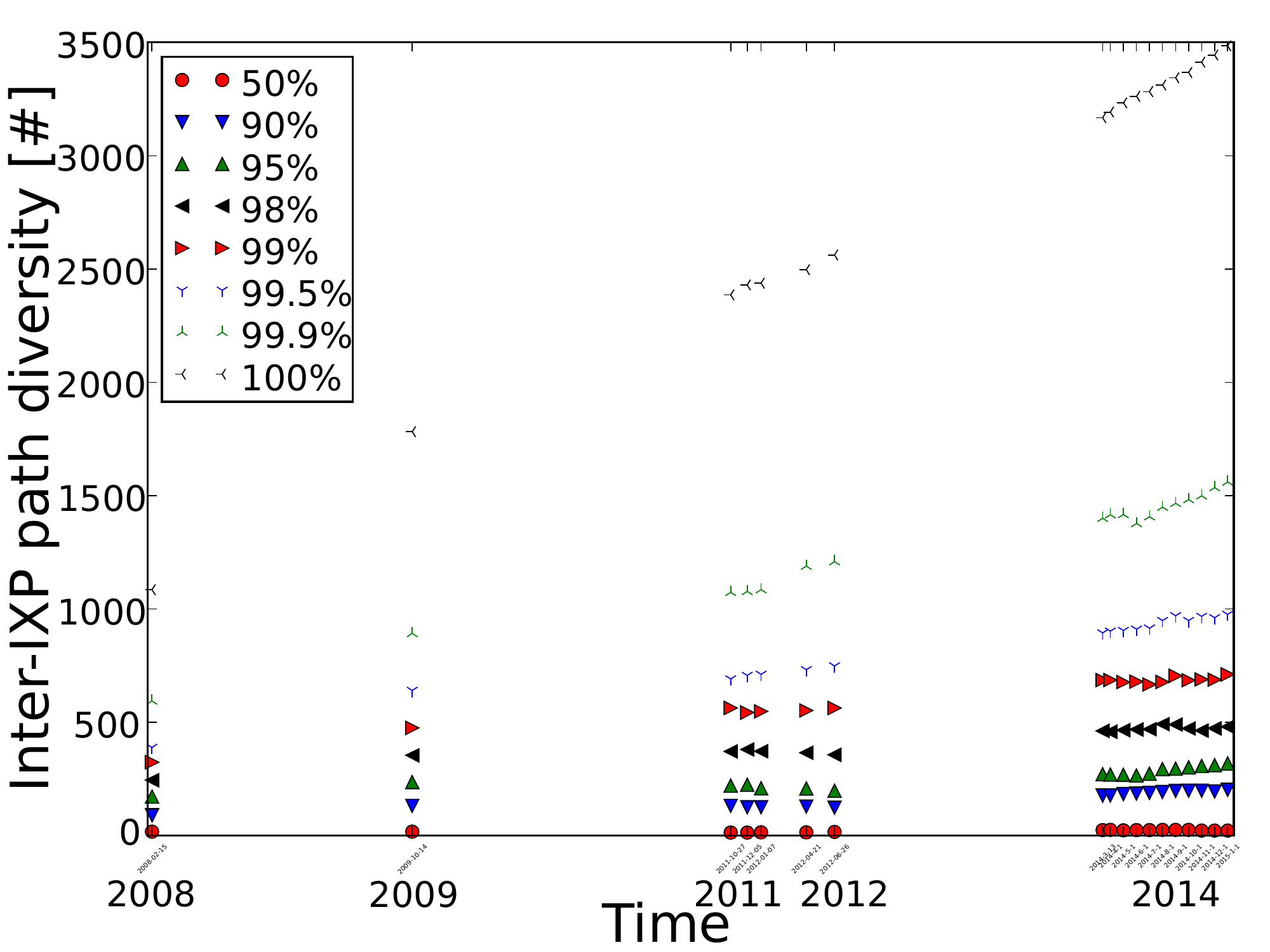}
    \caption{Edge-wise path diversity distribution percentiles betweeen all the candidate \emph{(IXP-IXP)} pairs.}
    \label{fig:pdb-path-div-perc}
  \end{subfigure}
  \caption{Edge multiplicity and edge-wise path diversity of the PeeringDB-based CXP multigraph over time.}
  \label{fig:pdb-ixp-edge-mult-path-div}
\end{figure}

In \fref{fig:pdb-path-div-perc}, we show the distribution percentiles 
of the path diversity between \emph{all} candidate IXP pairs. The diversity is calculated
as the number of edge-disjoint paths between each pair, computed with the minimum cut. 
We see that our observations regarding the edge multiplicity of \fref{fig:pdb-edge-mult-perc}
are amplified by about one order of magnitude. That is, the connectivity-wise rich IXP pairs 
compose a dense multigraph core, leading to a substantial 10-fold increase in the overall path diversity
as opposed to edge multiplicity. 

In summary, the IXP overlay graph is growing both in number of vertices and number of edges, thus improving connectivity. 
This is mainly due to the more aggressive peering of big players like Hurricane Electric, and the
introduction of many new IXPs in remote parts of the globe during recent years.~The edge multiplicity in the corresponding multigraph leads to an order of magnitude larger
path diversity over any IXP pair (with 1000s of paths available between the upper 0.1\% of the pairs).
This is intensified as time progresses, especially in the recent years. A heavy tail of well-connected IXPs
and aggressive AS peers is responsible for the dynamic expansion of the multigraph.
In the presence of this densely connected ``core", low path choice typically stems from badly connected stub ASes and not from a general graph property.

\makeatletter
\newcommand{\removelatexerror}{\let\@latex@error\@gobble}
\makeatother

\def\NULL/{\textnormal{\texttt{null}}}

\newcounter{ipCounter}
\NewDocumentEnvironment{IPFormulation}{m}{%
\refstepcounter{ipCounter}
\begin{algorithm}[#1]%
\renewcommand\thealgocf{\arabic{ipCounter}}
}{%
\end{algorithm}
\addtocounter{algocf}{-1}
}

\newcommand{\tagIt}{\refstepcounter{IPnumber}\tag{IP-\theIPnumber}}
\newcounter{IPnumber}
\setcounter{IPnumber}{0}

\newenvironment{ORIG}{\par\color{blue} \textcolor{red}{[BEGIN Vassilis' original]}}{\textcolor{red}{[END Vassilis' original]} \par}

\newenvironment{REWRITTEN}{\par\color{blue} \textcolor{red}{[BEGIN Vassilis' original: deprecated]}}{\textcolor{red}{[END Vassilis' original: deprecated]} \par}

\newtheorem{theorem}{Theorem}

\section{Path Stitching Algorithms}
\label{sec:algo}

As shown in \xref{sec:ixp-analysis} the IXP-based multigraph, on which
CXPs may operate, is very dense.
In this section, we present
algorithms to exploit its rich path diversity in order to maximize the number
of concurrently embeddable routes subject to QoS guarantees, such as maximal latency or minimal bandwidth.
These algorithms serve as the application logic of a logically centralized CXP controller, operating on
the global view of the IXP multigraph for inter-domain path stitching.

The problem that we need to solve is complex for
several reasons. \first Requests from the large client base (cf. \xref{sec:scaling}) dynamically arrive over time in a
non-predictable manner, necessitating the use of online algorithms. \second While
a single suitable e2e path can be found in polynomial time, the IXP-based graph offers rich choice (cf. \xref{sec:anatomy}, \xref{sec:policies}) and
requires to carefully select which of the edges between two IXPs to use.
\third The online selection of e2e paths should reflect multiple conflicting high-level
objectives, namely accepting as many requests as possible, avoiding the use of scarce low-latency, high-bandwidth edges, and preventing resource fragmentation.
We formally introduce the e2e routing problem considered in this work as the QoS Multigraph Routing
Problem (\textsc{QMRP}) in Section~\ref{sec:QoSPahtletRoutingProblem}, together with an optimal
offline formulation. Subsequently, we present a general algorithmic framework to solve
the \textsc{QMRP} in an online manner. In particular, given the computational complexity of the
problem, we employ a \emph{sample-select approach}, where in the first stage, a set of \emph{feasible}
paths is \emph{sampled}, and subsequently one of
them is \emph{selected} for the actual embedding (cf. Section~\ref{sec:online-sample-select}).
Lastly, the framework is extended to support reconfigurations of pre-generated embeddings
in order to accommodate further online requests.%in \xref{sec:adding-reconfiguration-support}.

\subsection{The QoS Multigraph Routing Problem}
\label{sec:QoSPahtletRoutingProblem}

We model the IXPs and their pathlet interconnections as a directed multigraph \(G = (\VG, \EG)\),
where $\VG$ is a set of IXPs (nodes/vertices) and $\EG$ are inter-IXP pathlets (links/edges) offered by ISPs. The ISPs annotate their
pathlets $e \in \EG$ with their available bandwidth
$\bE \in \mathbb{R}_{\geq 0}$ and their
latency $\lE \in \mathbb{R}_{\geq0}$.
On this substrate, we want to embed a set of e2e routing requests, henceforth denoted by \(\mathcal{R}\).
A request \(R\in \mathcal{R}\) asks for the establishment of an e2e connection between IP addresses $\sR$ and $\tR$
with minimal bandwidth $\bR$ and maximal latency $\lR$. Note that these start and end points are not included
in the pathlet network $G$.
However each IP address is, by its access ISP affiliation, implicitly connected to one or multiple IXPs (\cf \fref{fig:cxp-concept}).
While we take these multiple start and end IXPs into account in the
implementation of the presented algorithms, we assume simple IXP start and end points for the
sake of easier representation.

We study how CXP operators can accept (and embed) as many requests as possible---a natural objective for any
revenue-driven provider aiming at the maximization of its client base. Embedding a request $R \in \mathcal{R}$ here
refers to finding a suitable path $P_R$, such that the latency of $P_R$ is less than $\lR$ and that the
path $P_R$ can carry more than the minimal bandwidth $\bR$. Importantly, as inter-IXP pathlets can be used
by multiple requests, the maximal available bandwidth (\ie capacity) of pathlets must never be exceeded.

The offline version of the QoS Multigraph Routing Problem (\textsc{QMRP}), \ie when $\mathcal{R}$ is
given ahead of time, can be formulated as an Integer Program, \cf \emph{Integer Program~\ref{alg:IP}} (OptFlow):
the binary variable $x_R$ decides whether request $R \in \mathcal{R}$ is embedded and the variable $P^e_R$
indicates whether edge $e \in \EG$ is used by request $R \in \mathcal{R}$. The correctness of the formulation
stems from the following observations:
\first Constraints~\ref{alg:IP:induceFlow} and \ref{alg:IP:flowConservation} induce a unit flow from $\sR$ towards $\tR$ if
request $R \in \mathcal{R}$ is embedded (\cf \cite{ahuja1993network}). $\delta^+(v)$ and $\delta^-(v)$ here denote the
set of outgoing and incoming edges of $v \in V_G$ respectively.
\second By Constraint~\ref{alg:IP:latency} the path described by variables $P_R$ must obey the maximal latency $\lR$.
\third By Constraint~\ref{alg:IP:bandwidth} the available bandwidth (\ie capacity) of any pathlet is not exceeded.
While the offline problem is interesting for optimizing existing allocations of requests
in the background and further
increase acceptance ratios (see Section~\ref{sec:adding-reconfiguration-support}), we are in general
more interested in the online variant. In this context, each request $R$ is known only
at its arrival time, and the algorithm needs to compute an embedding (for the duration of the request)
at that time.

\renewcommand{\tagIt}{\refstepcounter{IPnumber}\tag{OF-\theIPnumber}}
\begin{figure}[t!]
\noindent
\scalebox{0.82}{
\begin{minipage}{1.2\columnwidth}
\removelatexerror% Nullify \@latex@error
\begin{IPFormulation}{H}
\removelatexerror% Nullify \@latex@error
\SetAlgorithmName{Integer Program}{}{{}}

\newcommand{\spaceIt}{\qquad\quad\quad}
\newcommand{\miniSpace}{\hspace{1.5pt}}

\BlankLine
\begin{fleqn}[0pt]
\begin{alignat}{3}
\phantomsection	   \textnormal{max~} & \sum_{R \in \mathcal{R}} x_R&  \tag{OBJ}  \\
\phantomsection
\label{alg:IP:induceFlow} x_R = & \sum_{e \in \delta^+(\sR)} P^e_R - \sum_{e \in \delta^-(\sR)} P^e_R & \forall R \in \Requests \tagIt \\
\label{alg:IP:flowConservation} 0 = & ~\sum_{e \in \delta^+(v)} P^e_R ~- \hspace{2pt}\sum_{e \in \delta^-(v)} P^e_R \quad & \vspace{-80pt}\begin{array}{r}
\forall R \in \Requests. \\
v \in \VG \setminus \{\sR, \tR\}
\end{array} \tagIt \\
\label{alg:IP:bandwidth} \phantomsection \bE \geq & \sum_{R \in \Requests} \bR \cdot P^e_R &  \forall~ e \in \EG \tagIt \\
\label{alg:IP:latency} \phantomsection  \lR \geq & \sum_{e \in \EG} \lE \cdot P^e_R &  \forall~ R \in \Requests \tagIt \\
\phantomsection x_R \in & \{0,1\} &\forall~ R \in \Requests  \tagIt \\
\phantomsection  P^e_R \in & \{0,1\} &\forall~ R \in \Requests, e \in \EG \tagIt
\end{alignat}
\vspace{-10pt}
\end{fleqn}
\caption{Optimal Flow Formulation (OptFlow)}
\label{alg:IP}

\end{IPFormulation}
\end{minipage}

}
\end{figure}

\newpage
\subsection{Online Sample-Select Strategy}
\label{sec:online-sample-select}

In order to tackle the online variant of the QMRP we propose a \emph{sample-select} approach.
In the first stage a set of feasible paths is sampled from the set of all feasible paths. In the
second stage one of these paths is selected for the embedding. We employ this
approach as computing the \emph{optimal} path under multiple objectives and constraints
is generally NP-hard~\cite{multi-survey}, while the algorithm might need to handle workloads
of tens or hundreds of requests per second.

While investigating multiple path sampling strategies, we consider only a single selection
strategy which aims at minimizing the utilization of the network (in order to provide room for
many requests) and to secondarily penalize use of high-bandwidth and low-latency edges (since they are more scarce). 
We note that determining the best selection strategy
is interesting in its own right, but lies outside the scope of this paper. The strategy used can be summarized as follows:
\first Strictly prefer paths with a smaller hop count.
\second Among paths with the same hop count, choose the one with the minimal inverse utility, computed edge-wise $\forall e \in P_R$:
\[
InvU(e) = \frac{\bE}{\min \limits_{e' \in E_G(u,v)} \bE[e']} \cdot \frac{\max \limits_{e' \in E_G(u,v) } \lE[e']}{\lE} / |E_G(u,v)| ~,
\]
where $E_G(u,v)$ denotes the set of
edges between nodes $u,v$.

\begin{figure}[t!]
\noindent
\scalebox{0.9}{
\begin{minipage}{1.09\columnwidth}
\removelatexerror% Nullify \@latex@error
\begin{algorithm}[H]
\SetAlgoNoEnd
\removelatexerror% Nullify \@latex@error

\SetKwInOut{INPUT}{Input}
\SetKwInOut{OUTPUT}{Output}
\SetKwFunction{FeasibleEdges}{feasEdges}
\SetKwFunction{propagate}{propagate}
\SetKwFunction{reconstruct}{reconstruct}
\SetKwFunction{selectEdge}{selectEdge}
\SetKwFunction{selectNextHop}{selectNextHop}
\SetKwFunction{nodeScore}{nodeScore}

\LinesNumbered
\DontPrintSemicolon

\SetAlgorithmName{Algorithm}{}{{}}

\INPUT{Network $G = (V_G,E_G,\bE,\lE)$, \\~Request $R = (\sR,\tR,\bR,\lR)$}
\OUTPUT{Path $P_R$ to connect $\sR$ to $\tR$ or \NULL/}

\BlankLine

\textbf{sample} set of \emph{feasible} paths $\mathcal{P}_R$\;
\eIf{$\mathcal{P}_R \neq \emptyset$}
{
	\textbf{select} \emph{best path} $P_R \in \mathcal{P}_R$ and \textbf{embed} $R$ accordingly
}
{
	\textbf{return} \NULL/
}
\caption{Outline of Online Sample-Selection Algorithm}
\label{alg:pathSamplingPathGeneration}
\end{algorithm}
\end{minipage}
}
\end{figure}

In the following
we focus on the path sampling strategy, and present three different
algorithmic variants.
The goal of the sampling algorithm is to efficiently compile a set of paths,
giving us the flexibility of choice.
In particular,  we exploit the fact that
computing feasible solutions is \emph{not} NP-hard:
\vspace{-5pt}
\begin{theorem}\label{thm:paths}
A feasible path for a given request $R$ can be computed in polynomial time.
\end{theorem}
\vspace{-5pt}
\emph{Proof.} The proof is constructive. We first prune all edges $e \in E_G$ whose bandwidth is not sufficient to
support the minimal bandwidth requirement $\bR$. Projecting the resulting multigraph onto a simple graph
by replacing each set of edges with the minimal latency edge of the set, the simple graph $G'$ is obtained.
We can now perform any polynomial shortest-path algorithm to obtain the path $P'_R \in G'$, if such a path
exists. If $\sum_{e \in P'_R} \lE \leq \lR$, a feasible path was constructed; otherwise no such path can
exist. Assume that this process would not find a feasible path even though such a
path $P \in G$ exists. By replacing each edge of $P$ with the minimal latency edge of the corresponding
multi-edge set, a feasible path in $G'$ is constructed, proving the theorem. $\blacksquare$

Theorem~\ref{thm:paths} is an important building block for all our path sampling algorithms,
as it allows us to: \first abort the generation of paths early using a single shortest path
computation, and \second devise path sampling algorithms that will \emph{always} return
feasible paths, if they exist (\cf Korkmaz \etal~\cite{multi-con}).
In \xref{sec:algo} of our accompanying technical report~\cite{kotronis2015invtech}, 
three such algorithms are presented. The \emph{Perturbed Dijkstra}
(\texttt{PD}) algorithm is essentially a $k$-shortest paths variant \cite{eppstein1998finding}, strictly
minimizing latency. The \emph{Guided Dijkstra} (\texttt{GD}) algorithm broadens the search space as
edge selection is latency-independent, and the \emph{Guided Random Walk} (\texttt{GW}) algorithm aims at
finding arbitrary feasible paths.
The run-time complexity of these algorithms is bounded by $\mathcal{O}\left( k \cdot (|E_G| + |V_G| \log |V_G|) \right)$, with 
$k$-many feasible paths to sample and $|V_G|$ nodes and $|E_G|$ edges to operate on.
The algorithms can be used for different path sampling cases, ranging from 
purely deterministic variants (\texttt{PD}), to semi-randomized (\texttt{GD}) 
and fully randomized variants (\texttt{GW}).

\subsection{Adding Reconfiguration Support}

\label{sec:adding-reconfiguration-support}

\begin{figure}[t!]
\noindent
\scalebox{0.9}{
\begin{minipage}{1.09\columnwidth}
\removelatexerror% Nullify \@latex@error
\begin{algorithm}[H]
\SetAlgoNoEnd
\removelatexerror% Nullify \@latex@error

\SetKwInOut{INPUT}{Input}
\SetKwInOut{OUTPUT}{Output}
\SetKwFunction{FeasibleEdges}{feasEdges}
\SetKwFunction{propagate}{propagate}
\SetKwFunction{reconstruct}{reconstruct}
\SetKwFunction{selectEdge}{selectEdge}
\SetKwFunction{selectNextHop}{selectNextHop}
\SetKwFunction{nodeScore}{nodeScore}

\LinesNumbered
\DontPrintSemicolon

\SetAlgorithmName{Algorithm}{}{{}}

\INPUT{Initially rejected request $R^-$,\\~Accepted requests $\mathcal{R}^+$ with path $P_R^+$ for $R^+ \in \mathcal{R}^+$}

\BlankLine

\textbf{sample} set of \emph{feasible} paths $\mathcal{P}_{R^-}$ for $R^-$ in the \emph{empty} graph\;
\If{$\mathcal{P}_R \neq \emptyset$}
{
	\textbf{compute} conflicting requests   $~~~~~~~~~~\mathcal{P}_{\mathsf{confl}} = \{R^+ \in \mathcal{R}^+ | \exists P_{R^-} \in \mathcal{P}_{R^-}, P_{R^+} \cap P_{R^-} \neq \emptyset \} $\;
	\textbf{try to (re-)embed}  $\mathcal{P}_{\mathsf{confl}} \cup \{R^-\}$ by an offline algorithm
}

\caption{Offline Reconfiguration Scheme}
\label{alg:reconfiguration}
\end{algorithm}

\end{minipage}
}
\end{figure}

\setcounter{IPnumber}{0}

\renewcommand{\tagIt}{\refstepcounter{IPnumber}\tag{HP-\theIPnumber}}

\begin{figure}[t!]
{
\centering\noindent
\scalebox{0.9}{
\begin{minipage}{1.09\columnwidth}
\removelatexerror% Nullify \@latex@error
\begin{IPFormulation}{H}
\removelatexerror% Nullify \@latex@error
\SetAlgorithmName{Integer Program}{}{{}}

\newcommand{\spaceIt}{\qquad\quad\quad}
\newcommand{\miniSpace}{\hspace{1.5pt}}

\BlankLine
\begin{fleqn}[0pt]
\begin{alignat}{3}
\phantomsection	\textnormal{max~} & \sum_{R \in \mathcal{R}} x_R &  \tag{OBJ}  \\
\phantomsection \label{alg:OptPaths:PathSelection}  x_R =& \sum_{ P_{R} \in \mathcal{P}_{R}} y_{P_{R}} & \forall R \in \Requests \tagIt \\
\phantomsection \label{alg:OptPaths:bandwidth} \bE \geq & \sum_{R \in \Requests, e \in P_{R}} \bR \cdot y_{R} &  \forall~ e \in \EG \tagIt \\
\phantomsection  x_R \in & \{0,1\} &\forall~ R \in \Requests  \tagIt \\
\phantomsection  y_{P_{R}} \in & \{0,1\} &\forall~ R \in \Requests, P_{R} \in \mathcal{P}_{R} \tagIt
\end{alignat}
\vspace{-10pt}
\end{fleqn}
\caption{Heuristic Path Formulation (HeurPaths)}
\label{alg:HeurPaths}
\end{IPFormulation}

\end{minipage}
}
}

\end{figure}

The sample-select scheme as presented in Algorithm~\ref{alg:pathSamplingPathGeneration}
can be used to find good embeddings of e2e path requests arriving one-by-one over time.
In particular, the algorithms try to embed each arriving request if this is possible, otherwise
they reject the request. However, greedily embedding one request after the other may not be
optimal over time, and sometimes, it may be worthwhile to reconfigure existing paths in order
to defragment the current allocation and make space for additional requests.
Thus in the following, we propose a hybrid online-offline scheme which performs exactly that:
requests which are arriving online over time are embedded using one of the sample-select
approaches described above. However, in addition, we run an offline optimization procedure
in the background: this reconfigures \emph{sets of paths} in order to improve
acceptance ratios further. Such reconfigurations may be the only possibility to accept a request.

We thus extend the sample-selection scheme depicted as
Algorithm~\ref{alg:pathSamplingPathGeneration} with the fallback scheme depicted as
Algorithm~\ref{alg:reconfiguration}. Given a just rejected request $R^-$,
feasible paths are first sampled in the empty network, \ie without any embedded requests.
If feasible paths exist and the request $R^-$ could in general be embedded, all requests
that conflict with any of the found paths are selected in Line~3. In Line~4 the algorithm
tries to reconfigure conflicting requests and embed $R^-$. Note that the reconfiguration
task corresponds to solving the offline \textsc{QMRP}, where \emph{all} given requests
\emph{must be embedded}. While generally the Integer Program~\ref{alg:IP} could be used
by requiring $x_R = 1$ for all $R \in \mathcal{R}^+ \cup \{R^-\}$, its run-time is prohibitive.
We have therefore developed Integer Program~\ref{alg:HeurPaths}, which does not compute paths on
its own, but is given the set of \emph{feasible} paths $\mathcal{P}_R$ of each request
$R \in \mathcal{R}$ as input, computed previously via online path-sampling. Again, by forcing $x_R = 1$ for all $R \in \mathcal{R}^+ \cup \{R^-\}$,
we can compute whether there exists a reconfiguration of embedded paths that allows accepting $R^-$,
increasing the overall acceptance ratio.

We note that the proposed formulation is an adaption of the classic multi-dimensional knapsack problem~\cite{freville2005multidimensional}
which, despite its NP-hardness, can be solved quite efficiently in practice using branch-and-bound
solvers~\cite{gurobi} when only dozens of paths are used for each request. In the evaluation (\cf~\xref{sec:eval}),
we use the \emph{HeurPaths} program as follows: first we produce sets of paths for all requests (5 to 20 per request)
using the previous path sampling algorithms on the initial empty graph, and then we employ \emph{HeurPaths}
to allocate the requests in an offline manner using the path set input.

\section{Algorithmic Evaluation}
\label{sec:eval}

We evaluate the performance of our algorithms in terms of Acceptance Ratio (AR),
utilization \ie the ratio of occupied bandwidth to the total
available capacity\footnote{This metric takes into account
only inter-IXP pathlets.}, and computation time per request.
To maximize the revenue, the acceptance ratio should be maximized while minimizing the resource utilization (so that there is room for more requests). Additionally, based on the ad-hoc online embedding of requests, the runtime should be low in order not to block the system. We use our custom CXP simulator\cite{cxp2015coderepo},
and the inter-IXP multigraphs described in \xref{sec:ixp-analysis}.
As discussed in Section~\ref{sec:algo}, based on the multigraph nature of the IXP graph 
existing algorithms are not suitable and need to be adapted accordingly.
In this section, we investigate the---empirical---trade-off between always choosing shortest paths (\emph{Perturbed Dijkstra}) versus 
the more randomized versions \emph{Guided Dijkstra} and \emph{Guided Random Walk} (cf. \cite{kotronis2015invtech}). Moreover, using the optimal Integer Program~\ref{alg:IP} as a baseline we show that our algorithms yield near optimal solutions quickly. 
Hence, our evaluation demonstrates how the orchestration on such graphs can be performed efficiently even on realistically sized scenarios. This is important for the application logic of potential SDN-based CXP implementations.
We next elaborate on the setup and main insights yielded by the evaluation process.

\begin{table}[t]
\centering
\scriptsize
\tabcolsep4pt
\begin{tabular}{ll|l}\toprule
\textbf{Parameter} & \textbf{Space (online)} & \textbf{Space (offline/hybrid)}\\\bottomrule
\multirow{4}{*}{Compared Algorithms} & Perturbed Dijkstra (PD) & HeurPaths-PD\\ & Guided Walk (GW) & HeurPaths-GW\\ & Guided Dijkstra (GD) & HeurPaths-GD\\ & & OptFlow\\\hline
Scaling-down factor (SDF) & 32, 16, 8, 4, 2, 1 & 32, 16\\\hline
Request Latency & \multicolumn{2}{c}{unif(100,150), (150,200), (200,250), (250,300) ms}\\\hline
Paths per request & \multicolumn{2}{c}{5, 10, 20}\\\hline
Number of requests per run & \multicolumn{2}{c}{10,000}\\\hline
\end{tabular}
\caption{Online and offline/hybrid parameter space.}
\label{tab:sim-params}
\end{table}

\begin{figure*}
  \centering

  \begin{subfigure}[b]{0.3\textwidth}
    \includegraphics[trim= 0mm 0mm 0mm 0mm, width=1.0\columnwidth]{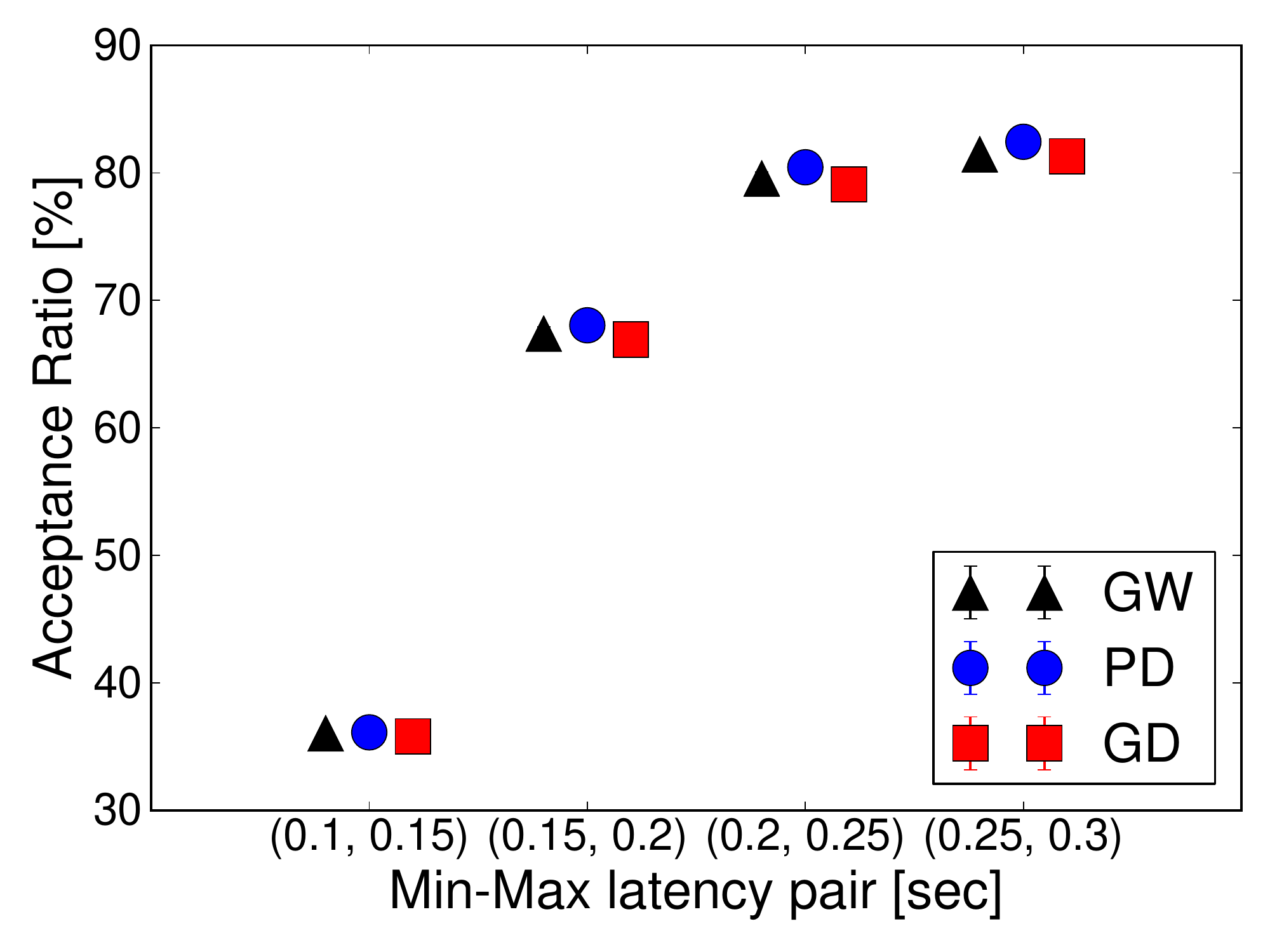}
    \caption{Acceptance Ratio vs Required Latency: SDF=8, 20 paths/request, 10,000 requests}
    \label{fig:ONLINE_AR_vs_lat}
  \end{subfigure}
  \hfill
  \begin{subfigure}[b]{0.3\textwidth}
    \includegraphics[trim= 0mm 0mm 0mm 0mm, width=1.0\columnwidth]{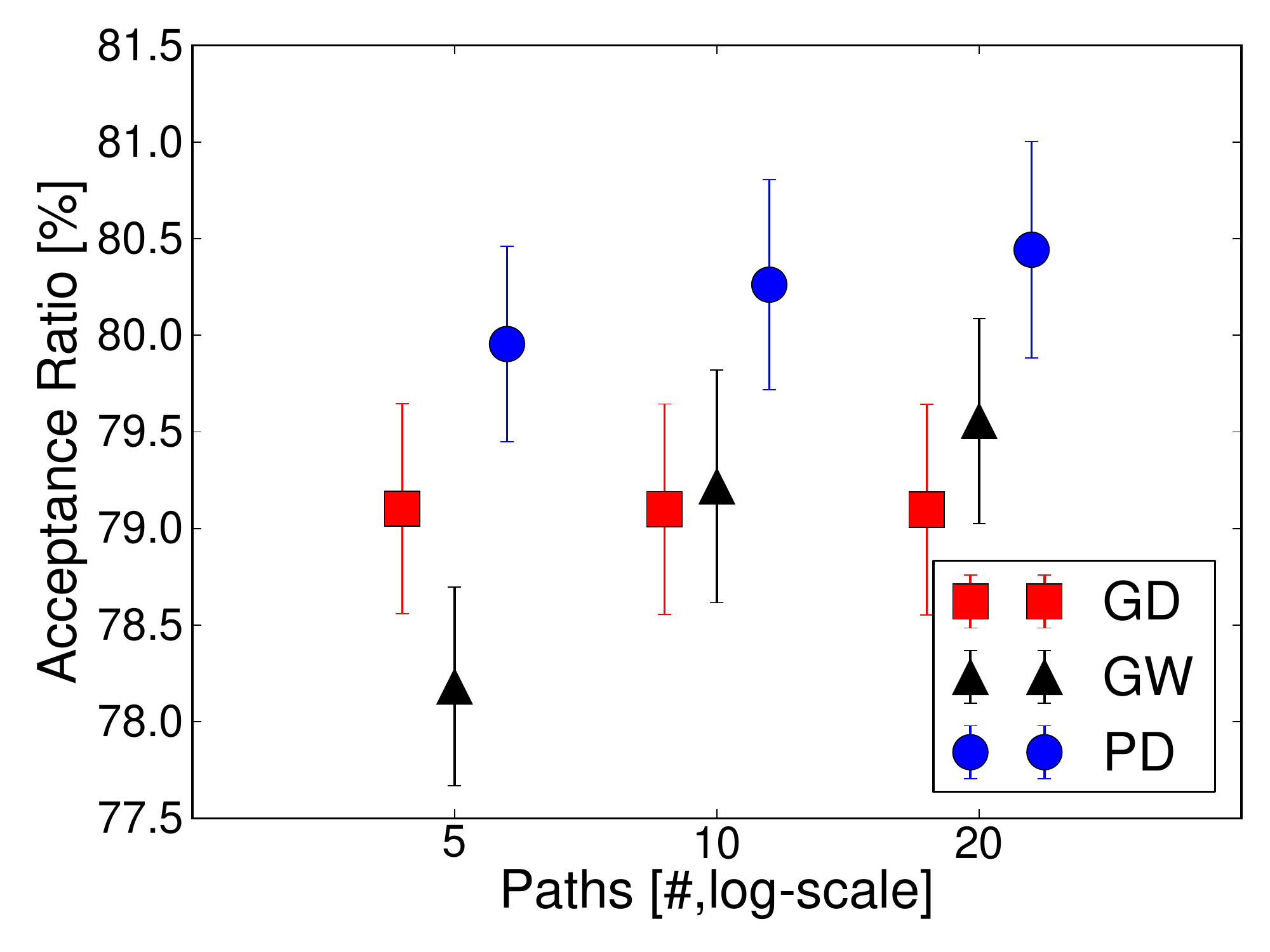}
    \caption{Acceptance Ratio vs paths/request: SDF=8, latency in (200,250) msec, 10,000 requests}
    \label{fig:ONLINE_AR_vs_np}
  \end{subfigure}
  \hfill
  \begin{subfigure}[b]{0.3\textwidth}
    \includegraphics[trim= 0mm 0mm 0mm 0mm, width=1.0\columnwidth]{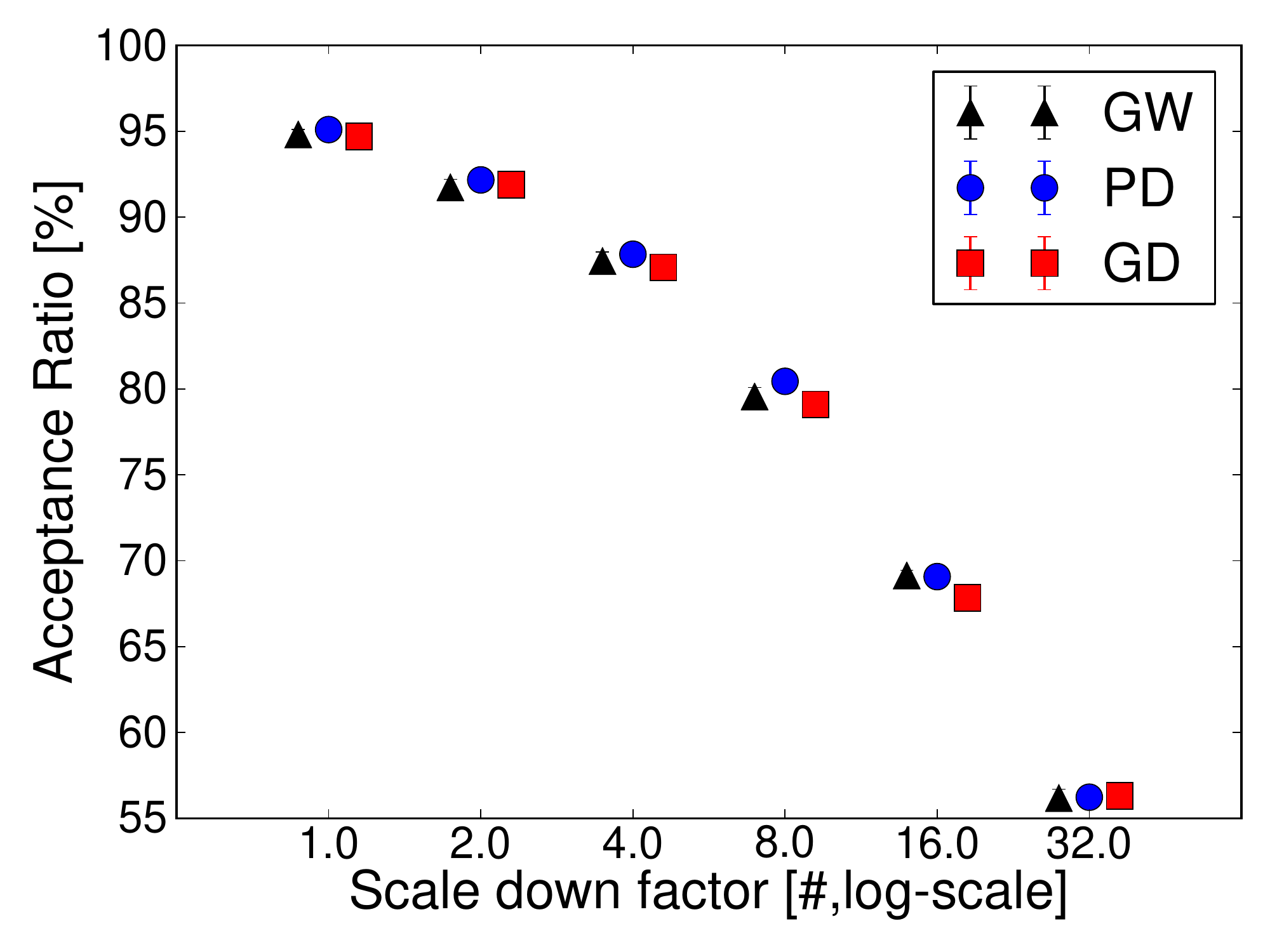}
    \caption{Acceptance Ratio vs SDF: 20 paths/request, latency in (200,250) msec, 10,000 requests}
    \label{fig:ONLINE_AR_vs_sdf}
  \end{subfigure}

  \begin{subfigure}[b]{0.3\textwidth}
    \includegraphics[trim= 0mm 0mm 0mm 0mm, width=1.0\columnwidth]{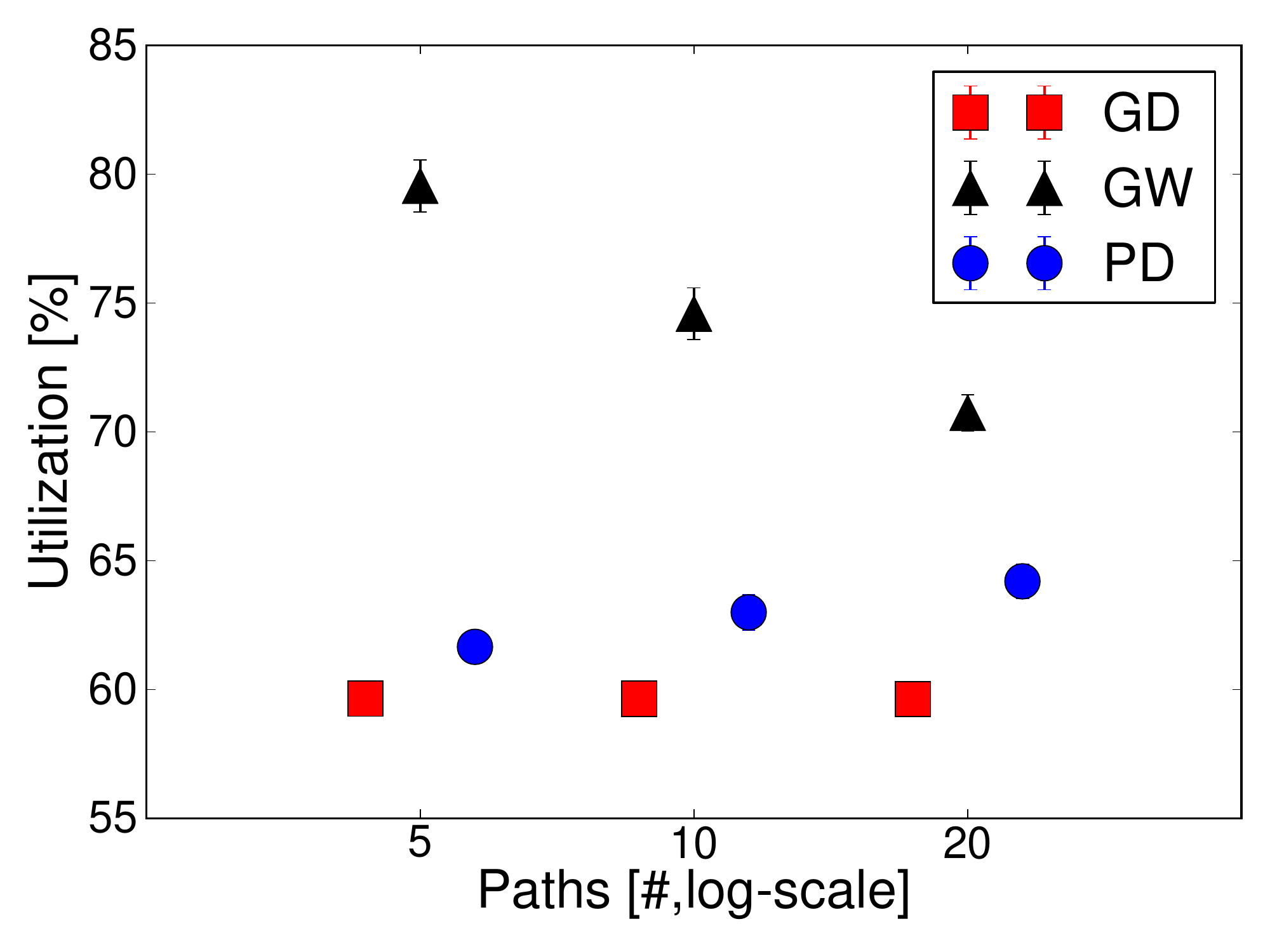}
    \caption{Utilization vs paths/request: SDF=8, latency in (200,250) msec, 10,000 requests}
    \label{fig:ONLINE_UTIL_vs_np}
  \end{subfigure}
  \hfill
  \begin{subfigure}[b]{0.3\textwidth}
    \includegraphics[trim= 0mm 0mm 0mm 0mm, width=1.0\columnwidth]{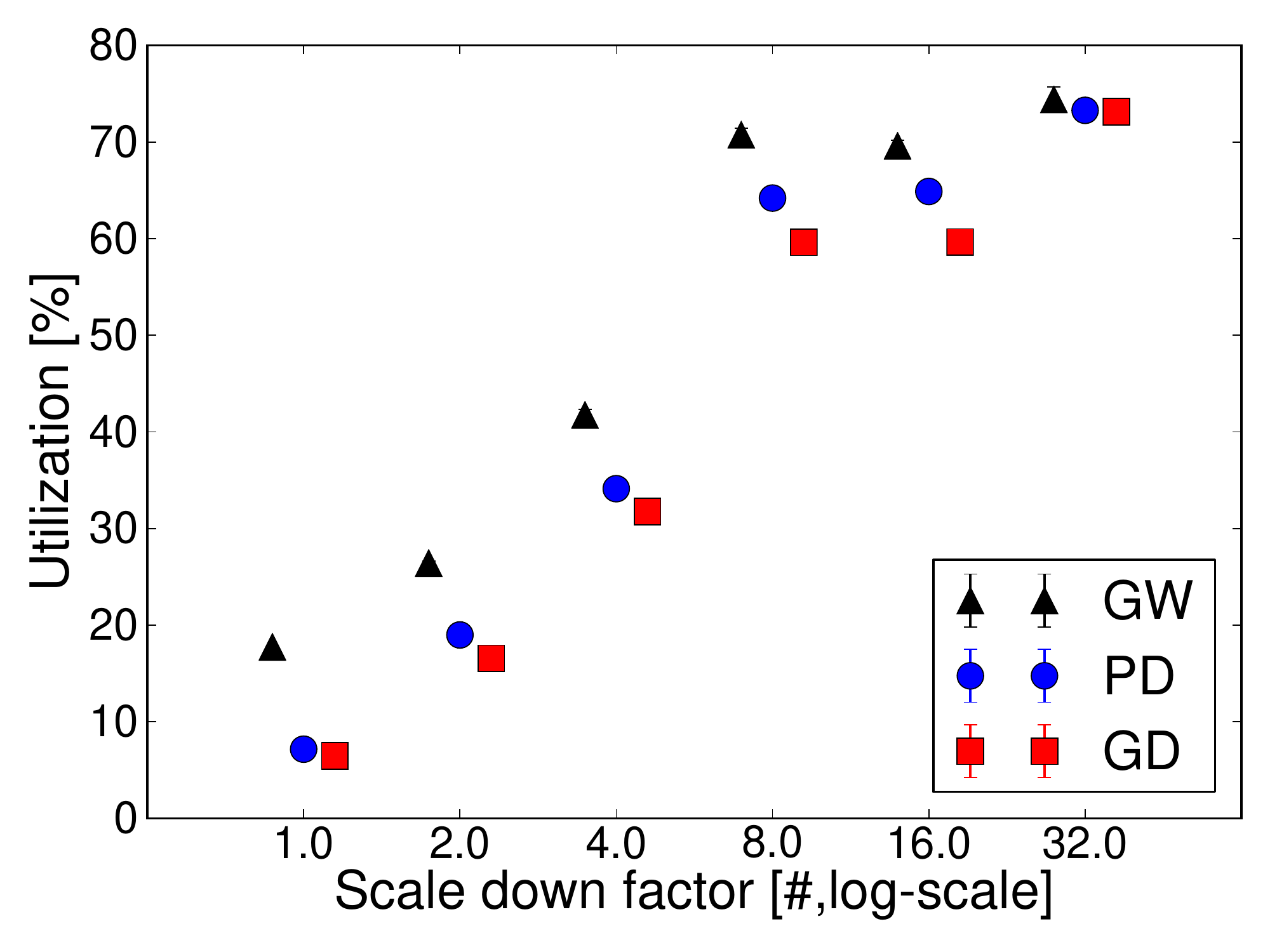}
    \caption{Utilization vs SDF: 20 paths/request, latency in (200,250) msec, 10,000 requests}
    \label{fig:ONLINE_UTIL_vs_sdf}
  \end{subfigure}
  \hfill
  \begin{subfigure}[b]{0.3\textwidth}
    \includegraphics[trim= 0mm 0mm 0mm 0mm, width=1.0\columnwidth]{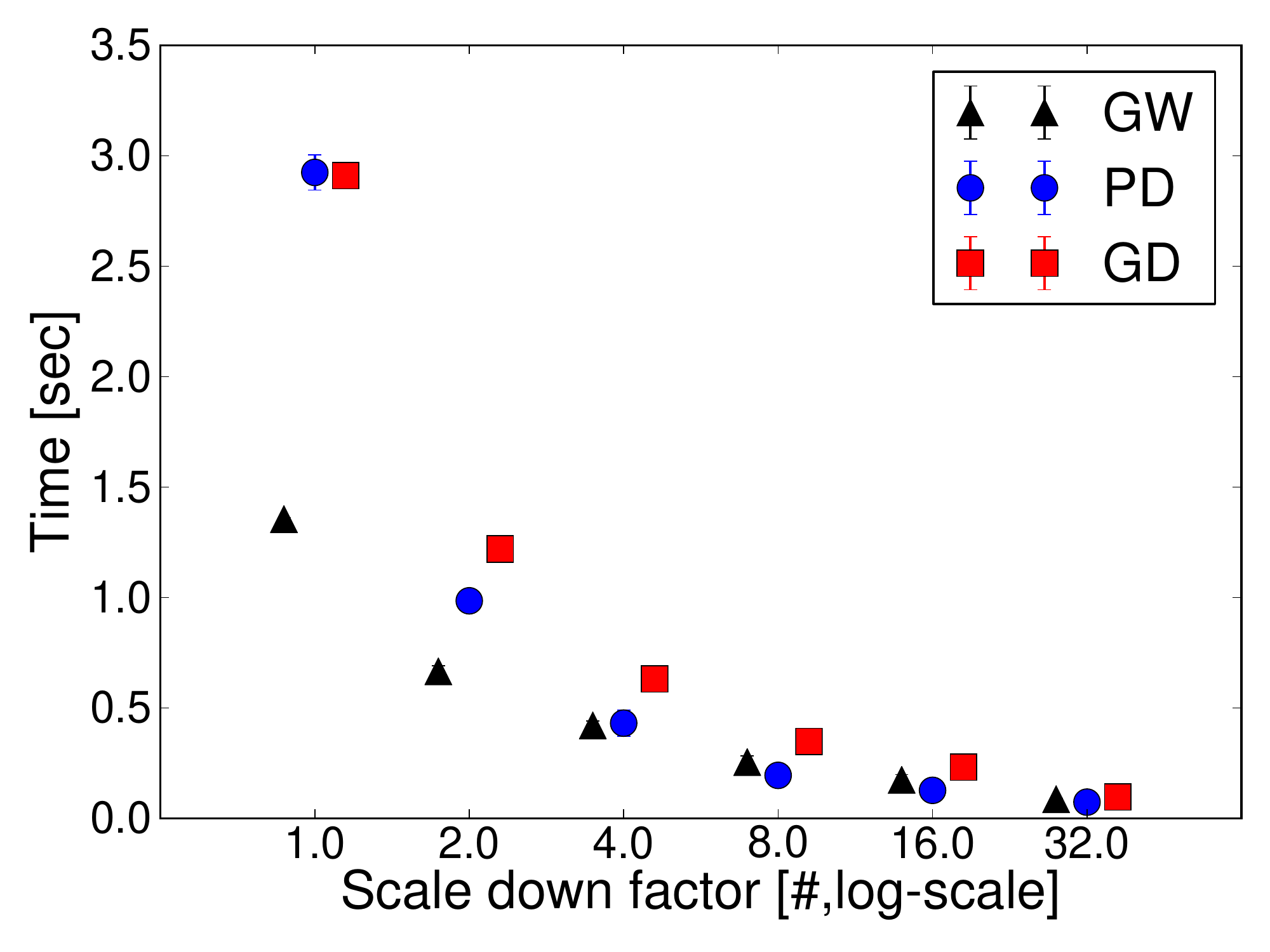}
    \caption{Time per request vs SDF: 20 paths/request, latency in (200,250) msec, 10,000 requests}
    \label{fig:ONLINE_TIME_vs_sdf}
  \end{subfigure}

  \caption{Moderate scale online simulation}
  \label{fig:mod-scale-online}
\end{figure*}

\subsection{Experimental Setup}

The search space of our simulations is composed of the cross-product
of the following parameter dimensions: \first pathlet latencies and \second bandwidths,
\third requested latencies and \fourth bandwidths,
\fifth graph sizes, \sixth maximal number of
paths generated per request, \seventh number of requests per simulation run,
and \eighth temporal characteristics of requests (\eg durations).
This search space has to be explored for each evaluated algorithmic variant.
Due to its large volume, we constrain our search space 
so that the simulations may run within reasonable time frames ($\sim$several weeks).
\tref{tab:sim-params} summarizes the used parameters. 
We next elaborate on the inter-IXP and endpoint-to-IXP pathlet latency
and bandwidth model, the choice of the request endpoints and the temporal
characteristics of the requests.

\textbf{Latency.} Pathlets connecting IXPs pass over ISP domains. To model pathlet latency in a
geographically diverse ISP, we utilize the Hurricane Electric (HE)
looking glass server~\cite{HELookingGlass}
and perform measurements between pairs of routers situated
at major PoPs around the world.
The variance of the measured latencies appears not to depend on the geographical distance $d$.
We therefore model the RTT as a linear function (parameterized by~$a$ and~$b$)
of $d$ combined with a random variable~$X$ to reflect the
uncertainty in the model:
$rtt(d) = a \cdot d + b + X$.
Through linear regression we find:
$a =  0.016 [\frac{\text{ms}}{\text{km}}]$
and $b =  26 \text{[ms]}$.
By least squares fitting, we model $X$ as a normal distribution
$N(\mu,\sigma)$ with $\mu = 0$ and $\sigma = 14 \text{[ms]}$.
We approximate the one-way latency as: $\lE = 1/2 \cdot rtt(d)$.
and use this model for both access and transit pathlets (\cf \xref{sec:use-case}).
Request latencies are selected uniformly at random from four ranges (\cf \tref{tab:sim-params})
to evaluate looser to stricter requirements.

\textbf{Bandwidth.}
We consider unitary requests, where each embedded request occupies the
full bandwidth of the edge(s) it uses. This simplification removes
the necessity to model offered and requested bandwidth and hence reduces the search space. 
In contrast, we rather focus on assessing our algorithms based on the topological
characteristics of the inter-IXP substrate. In particular, we ``fill'' the multigraph
with allocated bandwidth in order to discover its inherent potential for hosting
arbitrary requests. Moreover, the chosen setup of uniform (and thus blocking) paths gives 
insights in how well the choice of shortest paths with respect to the hop count and the 
actual latencies are balanced to achieve the best resource utilization. If one 
were to always prefer shorter paths with respect to the hop count, the resource footprint would be minimized; however, 
this would reduce the availability of ``mission-critical'' pathlets of very low latency, hence reducing the acceptance ratio for latency-sensitive requests in the long run. 
Non-unitary request settings and corresponding simulations and effects are the subject of future work.
For simplicity, non-access IXP-IXP pathlets are aligned with the unitary request bandwidth setting. In reality, their bandwidth is generally determined by ISP competition and auctioning~\cite{MINT}.

\textbf{Request Endpoints.}
We choose candidate IP addresses uniformly at random from the IPv4 address space adjacent
to the members of the IXPs under examination.
After we choose a source and destination address for a request, we retrieve
their respective coordinates using the MaxMind GeoIP2 database~\cite{GeoIP}.
These coordinates, together with the IXP locations, are used for geographical
distance calculations between endpoints and IXPs.
We assume that IP-IXP pathlets are
not constrained by bandwidth, since the access ISP can offer exactly the bandwidth requested in direct
collaboration with its client, even without CXP-based mediation.

\textbf{Online Requests.} The requests arrive in order
and are handled one-by-one in an online fashion.
Each embedded request persists during the lifetime of the simulation (``infinite'' duration),
so that the peak load in the online case corresponds to the offline
case, allowing for fair comparison at the corresponding graph scales.

\subsection{Observations \& Insights}

\fref{fig:mod-scale-online} and~\fref{fig:small-scale-offline} present
key observations regarding algorithmic performance, which we further
explain and analyze below. Note
that the ranges on the y-axes do not have a 0-baseline but are adapted
per figure. All results are based on 10 runs per simulation.
We show average values with error-bars of 1 standard deviation. The baseline
algorithm for the online case is the \emph{Perturbed Dijkstra}, while \emph{OptFlow}
is the offline/hybrid baseline variant.
We note that simple-graph approaches and baselines of previous work, not tailored to multigraph sampling (\cf \xref{sec:related}), would have
to operate on orders of magnitude larger substrates, \eg using 2 ``half-edges'' and one AS node to simulate a pathlet, inducing biases.
In such graphs $|\VG| = O(|\EG|)$, while here $|\VG|\ll|\EG|$.

\textbf{Which online path sampling algorithm allows for the maximal acceptance
ratio, at the lowest utilization?}
The winner in terms of acceptance ratio is the Perturbed Dijkstra
approach with a lead of 1-2\% (\cf \fref{fig:ONLINE_AR_vs_lat}, \fref{fig:ONLINE_AR_vs_np},
\fref{fig:ONLINE_AR_vs_sdf}), as opposed to Guided Dijkstra and Guided Walk. In terms of utilization,
Guided Dijkstra wins by about 2-5\% followed closely by Perturbed Dijkstra, while the Guided Walk is worse
within a best-case gap of about 10\% from its Dijkstra-based counterparts (\cf \fref{fig:ONLINE_UTIL_vs_np}),
across scales (\cf \fref{fig:ONLINE_UTIL_vs_sdf}).
The reason for the prevalence of Perturbed Dijkstra regarding acceptance ratios lies in its $k$-shortest
path discovery; the edge-disjointness perturbation criterion, accompanied by the path selection
function (\cf \xref{sec:online-sample-select}), counteracts its tendency to consume precious (latency-wise) paths and leads to good embeddings.
Both Dijkstra approaches embed low-latency, low-hop paths that consume small amounts of bandwidth
on the substrate network. Especially the Guided Dijkstra performs shortest path routing on random
samples of the network, further lowering utilization. In contrast, the Guided Walk, due to the
fully randomized path sampling process, embeds feasible but higher-hop paths with an important penalty on utilization and
a small disadvantage in acceptance ratios. Its behavior in these two areas gets better as the number of
calculated paths increases (\cf \fref{fig:ONLINE_AR_vs_np}, \fref{fig:ONLINE_UTIL_vs_np}), since its progressive,
random path sampling process benefits from exploring richer path sets (\cf \xref{sec:online-sample-select}).

\textbf{How do hybrid variants behave w.r.t. acceptance ratios?}
HeurPaths with Guided Walk performs
the best in terms of acceptance ratios and is very close to the offline optimal values. In contrast, HeurPaths with
Perturbed or Guided Dijkstra leads to lower acceptance ratios as seen in
\fref{fig:OFFLINE_AR_vs_lat}, with differences up to 10\% for relaxed latency requirements.
This is explained with the optimal latency seeking stages of these algorithms that do not
couple well with the heuristic hybrid allocation. Thus they fail to exploit
the richness of the substrate, being biased towards the same low-latency edges. This
leads HeurPaths to saturation and limits maneuverability in path allocation. The advantage of
Guided Walk is preserved across scales (\cf \fref{fig:OFFLINE_AR_vs_sdf}) and latencies (\cf \fref{fig:OFFLINE_AR_vs_lat}).

\textbf{How do offline, hybrid and online algorithms compare with each other w.r.t. acceptance ratios and utilization?}
Our experiments on the 32-SDF and 16-SDF graphs show that the
online algorithms perform as good as the optimal offline and hybrid in terms of
acceptance ratios, but have 20-30\% lower utilization. The main
reason for this is the path selection criterion for the online
simulation (\cf \xref{sec:online-sample-select}), which prefers low-hop paths: the
online variants hit the optimal value through low utilization, while the offline variants
optimize based on sophisticated but computationally expensive
allocation of requests, ignoring utilization. Note that with SDFs of
32 and 16 due to the small number of IXPs and the nature of the
request model, many of the requests can be served directly using their
access ISPs and a single IXP, without occupying bandwidth on
the inter-IXP graph. We did not include larger graph sizes for OptFlow due to run-time scaling issues,
which we explain in the following.

\begin{figure*}
  \centering

  \begin{subfigure}[b]{0.3\textwidth}
    \includegraphics[trim= 0mm 0mm 0mm 0mm, width=1.0\columnwidth]{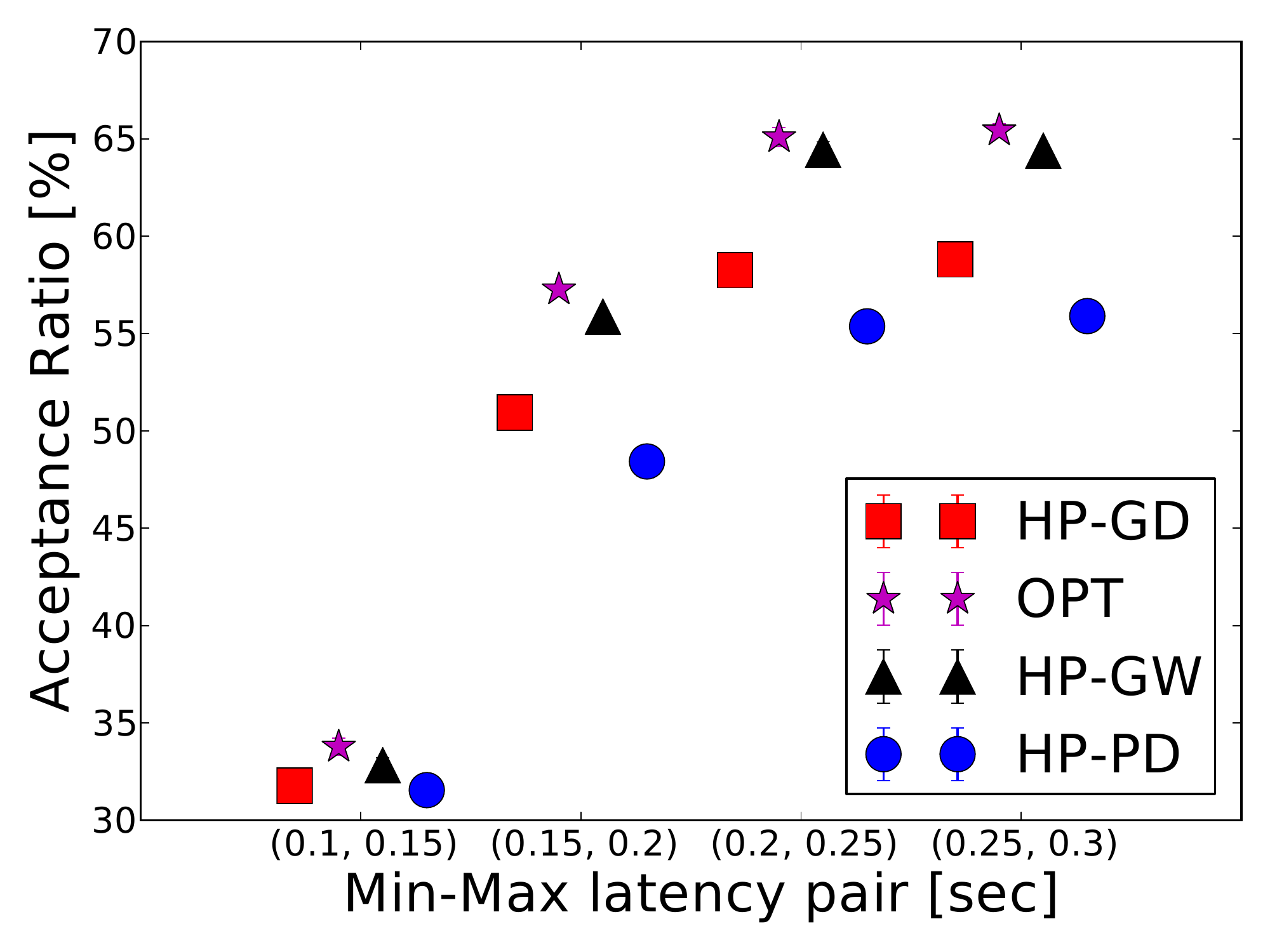}
    \caption{Acceptance Ratio vs Required Latency: SDF=16, 20 paths/request, 10,000 requests}
    \label{fig:OFFLINE_AR_vs_lat}
  \end{subfigure}
  \hfill
  \begin{subfigure}[b]{0.3\textwidth}
    \includegraphics[trim= 0mm 0mm 0mm 0mm, width=1.0\columnwidth]{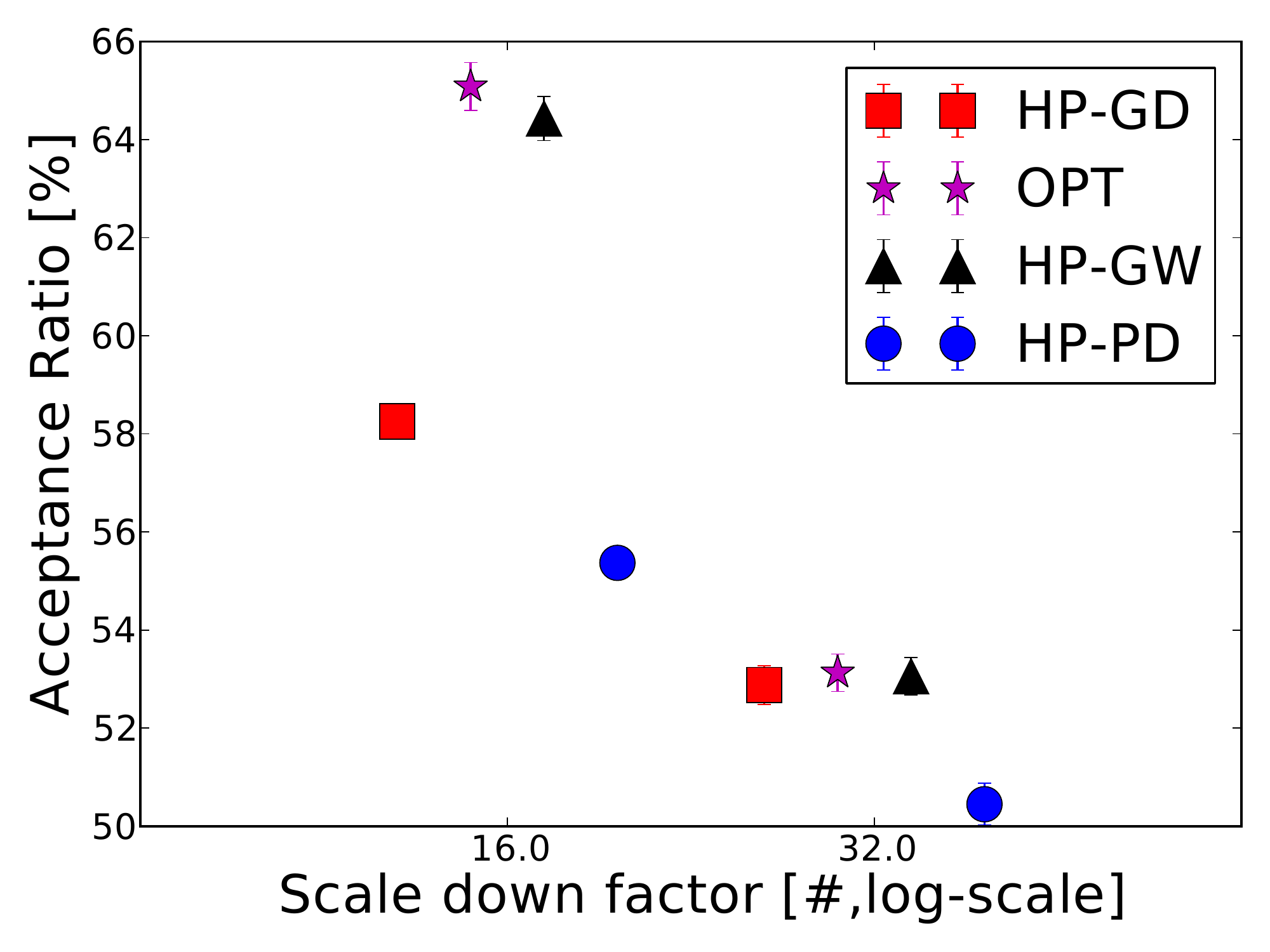}
    \caption{Acceptance Ratio vs SDF: 20 paths/request, latency in (200,250) msec, 10,000 requests}
    \label{fig:OFFLINE_AR_vs_sdf}
  \end{subfigure}
  \hfill
  \begin{subfigure}[b]{0.3\textwidth}
   \includegraphics[trim= 0mm 0mm 0mm 0mm, width=1.0\columnwidth]{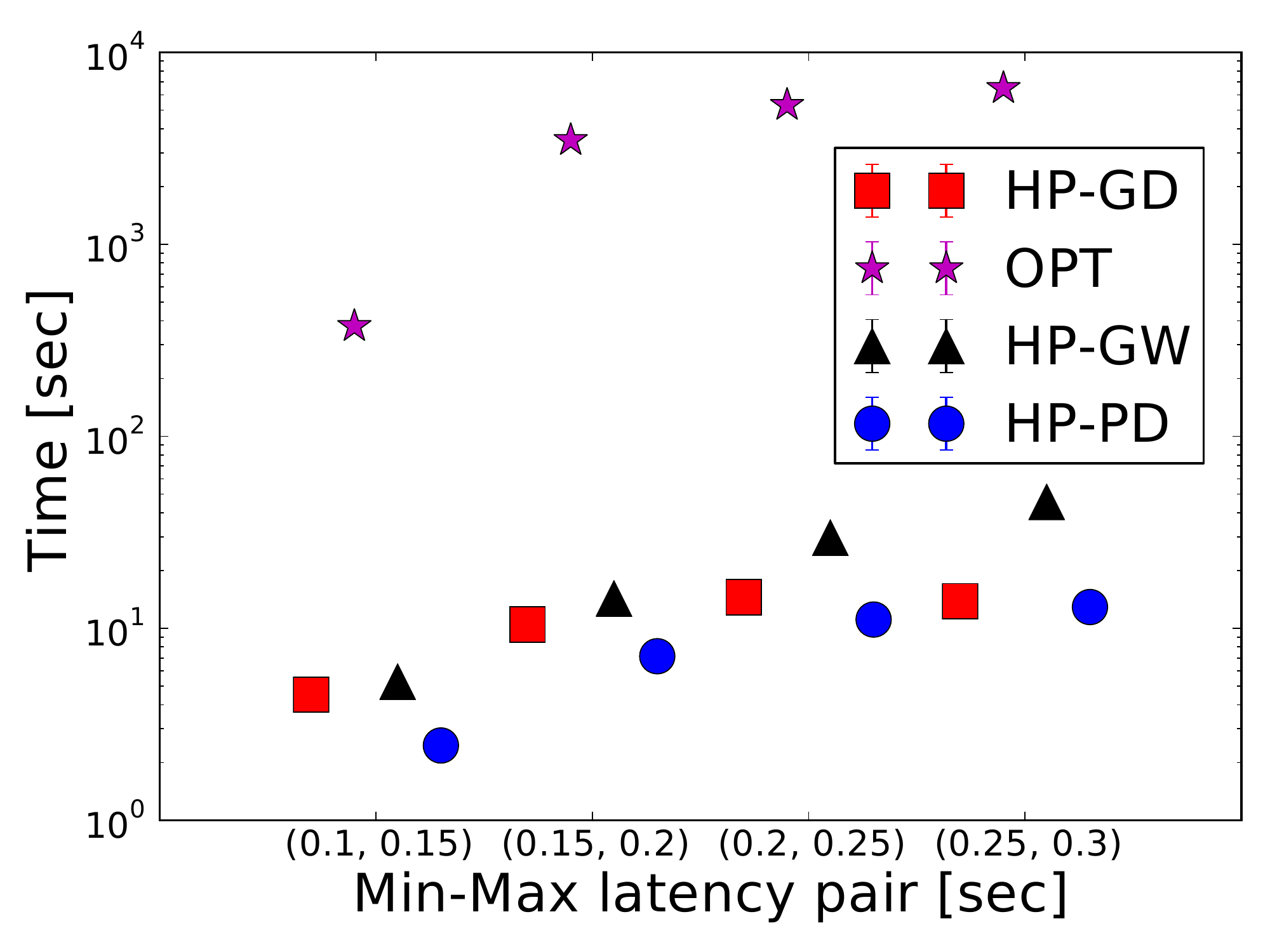}
    \caption{Mixed Integer Progam Time vs Latency: SDF=16, 20 paths/request, 10,000 requests}
    \label{fig:OFFLINE_TIME_vs_lat}
  \end{subfigure}

    \caption{Small scale offline/hybrid simulation}
  \label{fig:small-scale-offline}
\end{figure*}

\textbf{How do graph sizes affect run-times?}
Increasing the graph size (\ie lowering the SDF) leads to longer
run-times as expected, with the online Perturbed and Guided Dijkstras
scaling worse than the Guided Walk (\cf
\fref{fig:ONLINE_TIME_vs_sdf}). This is because the Guided Walk simply
finds \emph{feasible} paths quickly, without taking latency
\emph{optimality} into consideration and has lower computational
complexity (\cf \xref{sec:algo}). In contrast, the
optimal offline algorithm operates roughly at 1 to 3 orders of
magnitude slower than the hybrid variants at scales of 32-SDF or
16-SDF (\cf \fref{fig:OFFLINE_TIME_vs_lat}), and scales very poorly
for larger graphs. For the heuristic hybrid algorithm (HeurPaths) the
bottleneck is the preemptive path sampling for all requests, while
the path embedding stage has negligible time overhead.
The use of HeurPaths in collaboration with the Guided Walk yields
near-optimal acceptance ratios (\cf
\fref{fig:OFFLINE_AR_vs_lat},~\fref{fig:OFFLINE_AR_vs_sdf}) at
efficient run-times; the latter is evident in
\fref{fig:OFFLINE_TIME_vs_lat}, which presents the run-time of the
Mixed Integer Programming computations versus the requested latencies.
HeurPaths needs 10-100s to embed 10,000 paths. The path
computations can be parallelized, or be augmented by existing
online paths. For example, the Guided Dijkstra and Walk can be
parallelized after their first Dijkstra iteration, reducing run-times on
multiple cores.

\textbf{What is the effect of looser latency guarantees?}
The acceptance ratio (\cf
\fref{fig:ONLINE_AR_vs_lat},~\fref{fig:OFFLINE_AR_vs_lat}) and
utilization generally increase monotonically as the latency
requirements become looser, \ie less strict. This behavior comes to a halt when the
substrate is heavily utilized. The utilization ceiling is first hit by
the Guided Walk, then by the Perturbed Dijkstra and last by the
Guided Dijkstra. In the hybrid case increased latencies
and therefore increased search spaces widen the gap between
HeurPaths with Guided Walk and the Dijkstra-based variants as we can observe in
\fref{fig:OFFLINE_AR_vs_lat}.

\textbf{What have we learned from online-offline cooperation?}
We have observed that
using direct online-offline cooperation as described
in Algorithm~\ref{alg:reconfiguration} increases
acceptance ratios marginally ($\sim$1\%)
in overloaded (>70\%) substrates.
An interesting observation here relates to the request load
distribution. The optimal and heuristic offline algorithms have
increased utilization (20-30\% more than
the online variants), and do not improve too much in terms of
acceptance ratios when coupled with online request management. These variants solve the
problem purely from the perspective of maximizing the acceptance ratio for the \emph{entire current} set of
requests at their disposal, but have no incentive to optimize for utilization at the
same time. Thus they prefer to embed as many requests as possible, even at the cost of saturating
the substrate.
In contrast, the pure online variants cannot see all the requests concurrently;
therefore, they are doing their best to allocate each incoming request, or reject it when needed, without
sacrificing utilization and jeopardizing future acceptance.
We note that, depending on the CXP operator's goals, the heuristic
hybrid variant can be reformed to
optimize also for utilization and not only acceptance ratios, in order to efficiently
defragment the substrate when required.

\textbf{Summary: which algorithm should we prefer?}
In our experiments, we observed different behaviors in terms of
acceptance ratios in the online and hybrid case. In the online case,
Dijkstra-based approaches prevail, while in the hybrid case fully
randomized sampling performs better. More precisely, in the online
scenario Perturbed Dijkstra is a better choice at small graph scales
because of its high acceptance ratios and low utilization; at these
scales the run-time of all algorithms is short. We would opt for
Guided Walks at large scales, when fast request allocation is
desirable, especially if the incoming load of requests is high (\eg
due to higher CXP penetration). In this case, rich path sets (\eg
20 per request) are important, since they allow the Guided Walk to
achieve good acceptance ratios at reasonable utilization levels, which are
close to its Dijkstra-based counterparts.
Lastly, HeurPaths is a much better candidate for scaling up the hybrid version
of the problem as opposed to OptFlow because it achieves similar
acceptance ratios---in particular when combined with the Guided Walk---at much shorter run-times.

\section{Telesurgery as a Use Case}
\label{sec:disc}

To get a better understanding of how CXPs can be used, beyond as plain multi-domain
bandwidth brokers, we investigate the telesurgery~\cite{telesurgery,marescaux2002transcontinental}
use case. Telesurgery undoubtedly has stringent
requirements on both availability and latency. Availability is
essential for ensuring uninterrupted surgical operations and the patient's
safety. Latency is important for making remote
surgery feasible with real-time feedback~\cite{telesurgery}. In addition, the bandwidth
requirements are generally high, \eg for transmission of video
streams~\cite{marescaux2002transcontinental}.

Regarding availability, quick fail-over in case of emergencies is challenging~\cite{4050103,QoSIPMPLS}.
As a consequence, higher redundancy is needed \emph{a priori} to
achieve acceptable availability. One way to achieve this
with CXPs, is to allocate multiple disjoint paths on the multigraph and send
redundant packet copies on each path. One copy is then selected by the receiver and 
delivered to the application. A more
efficient approach could be using Forward Error Correction (FEC) such as 
Reed-Solomon. For example, a CXP could allocate
12 disjoint paths with 1/10 of the required capacity each; then use a FEC
scheme with 12 channels including 2 times redundancy at 20\perc bandwidth overhead.
A CXP can check online for path failures. If a path is degraded, the CXP immediately allocates
a replacement, leaving the rest of the operational paths intact.
Obviously, less reliable paths within ISPs mandate more redundancy to achieve high availability.

In a CXP context, the ISP's network resources are virtualized. 
This example demonstrates how \emph{on-demand} resource provisioning may be used to bring prices down, 
by bringing up the utilization of the resource and amortizing its costs, analogously to 
how CPU and storage are better utilized in the context of cloud computing. 
Client flows can be dynamically assigned to (multiple) pathlets depending on the resources that are
available within the ``CXP cloud''.

CXPs may also be able to find lower latency
paths than traditional routing. If a path is subject to a triangle
inequality~\cite{lumezanu2009triangle} violation (the majority of paths
are~\cite{HotInterconnects}) and there is a well-placed CXP anchor
available to route over, the CXP can provide a path with
lower latency. This implies the need for a broad CXP deployment
footprint. While starting with selected IXPs as CXP anchors can serve as an
initial step, 
it may not be sufficient for optimizing latency~\cite{6423683}.

Finally, we note the following challenges related to SDN-based 
CXP implementations in the context of such use cases. 
\first Controller distribution and placement; the
RTT between the data plane anchors and the centralized CXP controllers
is a lower bound of the reaction times to failures or state updates, 
while full distribution induces state consistency and concurrency challenges~\cite{LogCen}.
\second Forming an accurate, real-time monitoring infrastructure for
supervising pathlet guarantees and measuring the performance of QoS-constrained flows is a
challenging task in its own right~\cite{QoSMon}. Nevertheless, CXPs 
need to control just a handful of IXP anchors around the world, which is a promising starting point.
Also, the complexity of pathlet formation and state monitoring is delegated to the ISP. 
For example, physical link failures that affect pathlets are first handled locally within the ISP and then propagate on the inter-domain level only if the failure needs to be known to the CXP to be handled via e2e 
re-routing.
\third Path embeddings need to be protected against failures via CXP
controller and anchor redundancy. These challenges are interesting directions
for future SDN research at the inter-domain level, in the context of centralized
pathlet stitching as a novel multi-IXP service.

\balance

\section{Related Work}
\label{sec:related}

\textbf{Internet QoS and Our Work.} Quality-of-Service is an evergreen topic that has been
discussed for decades~\cite{QoSPic,QoSServ,QoSDiff}, together with
the challenges associated with its implementation~\cite{QoSFail,QoSChallenges,4050103}.
Such challenges have hindered its Internet-scale adoption in parallel with classic
best-effort IP routing and peering agreements~\cite{IntScaleQoS}. Our work is 
complementary to existing work on end-to-end Internet QoS, which covers the spectrum
from low-level implementation (queueing mechanisms, QoS-oriented MPLS, OpenFlow mechanisms) to high-level
policies (SLAs, traffic isolation). We propose an IXP-centric model that can 
be used to support the deployment of inter-domain QoS in the context of centralized 
pathlet brokers and resource controllers (\cf \xref{sec:brokers}). CXPs could 
capitalize on prior work for the implementation~\cite{MPLSQoS,UniMulQoS,DistQoS,OpenQoS} and
monitoring~\cite{QoSMon} of QoS-enabled pathlets; the scheme assumes
a given per-ISP QoS and focuses on what can be done assuming that ISPs provide 
guaranteed pathlets anchored to IXPs, irrespective of \emph{how} they are implemented.
We note that this work, based on the concept of logically centralized brokers offering Routing as a Service~\cite{RaaS},
is an alternative to the proposal of source-routed, composable path segments advocated \eg in ARROW~\cite{peter2014one}.
We believe though that using IXPs as the primary points where the path composition takes place
could be common ground for the deployment of such proposals. Moreover, 
the CXP business model, involving the mediation of contracts between end-clients and pathlet providers, could benefit from works that facilitate the formation, establishment, and verification of end-to-end connectivity agreements based on cryptocurreny systems~\cite{castroroute}.

\textbf{IXPs.} Recently, a number of studies analyzed the important
role of IXPs~\cite{IXPsEye} in terms of: 
\first the flattening of the Internet
topology~\cite{FlatInternet,IXPStructure}, \second the prevalence of
IXP-based peering links in the Internet
ecosystem~\cite{ixp-anatomy,IXPMap}, and \third performance improvements,
such as the reduction of average Internet delays and path lengths~\cite{IXPsInternetDelays}.
The potential rise of SDN within IXPs, \eg enabled by
Software Defined Internet eXchanges (SDX)~\cite{SDX-SIGCOMM}, 
coupled with the changing role of IXPs, could turn to be an 
avenue for inter-domain QoS services based on the CXP paradigm.
Moreover, Hu \etal~\cite{6233057} investigated how a version of \emph{on-demand} peering policy relaxation
can take place at IXPs in order to recover from route failures. Our more general approach (\cf \xref{sec:ixp-analysis}) actively
uses the path diversity induced from different variants of routing
policies, based on sequential composition of \emph{inter-IXP} pathlets.
Finally, we refer the reader to the work of Castro \etal~\cite{castro2014remote} on remote peering at IXPs.
Among other things pertaining to their
study, the authors discuss the marginal utility of reaching extra ISPs in terms of the potential for offloading transit
traffic. In contrast, we are investigating the incremental deployment of IXP-based penetration
in terms of: \first end-client coverage, and \second path diversity potential for connecting these end-clients
under certain quality guarantees. However, the proliferation of remote peering practices means that the IXP-based multigraph
tends to get even richer, with remote ISPs being able to offer pathlets (using other layer-2 resellers) to more IXPs than the
ones in their direct vicinity.

\textbf{QoS Routing and Embeddings.}
Finding suitable paths between a pair of endpoints is a classic problem in computer science,
and has been studied intensively in the context of online call control~\cite{online-call},
virtual-circuit routing~\cite{circuit-routing, virtu-circuit} and also specifically
QoS provisioning~\cite{multi-survey}.
In the area of QoS routing, exact, approximate and heuristic algorithms have been
considered for finding paths subject to (possibly) multiple constraints and
objectives. Based on the dense nature of the CXP multigraph
and the online fashion in which requests arrive, we have adapted two well-known
heuristic algorithms frequently used in the context of QoS: k-shortest paths~\cite{eppstein1998finding,multi-survey}
and the look-ahead scheme employed by Korkmaz et al.~\cite{multi-con}.
In contrast to stochastic QoS routing algorithms as presented by Orda~\cite{Orda1999}, we assume QoS guarantees
over the provided ISP pathlets. Optimal solutions to the QoS routing problem are
generally NP-hard to achieve, due to having to consider multiple objectives (minimizing costs, avoiding scarce low-latency
links etc.) or multiple constraints (latency, bandwidth, jitter etc.)~\cite{multi-survey}.
The heuristic offline variant of our problem (embed as many e2e paths as possible), is a variant of
\emph{unsplittable} flow problems~\cite{Goemans-split} and is related to the VPN~\cite{vpn-embed} and virtual
testbed mapping~\cite{polyvine} problems. For a good survey, we refer the reader to Fischer \etal~\cite{emb-survey}.
Schaffrath \etal~\cite{VirtArch} also present a relevant virtualization architecture.
The \emph{hybrid online-offline} approach that enables the reconfiguration
of existing e2e embeddings, was shown to increase acceptance ratios in the domain of
virtual network embeddings by Fan and Ammar~\cite{fanAmmar}.
Frikha and Lahoud have recently proposed to precompute QoS paths to improve
performance~\cite{precomp}. In contrast, the paths that have already been computed in our work are reused
at a later stage (possibly in different contexts), thereby not introducing any additional computational overhead.
Lastly, Ascigil \etal~\cite{ascigil2014scalability} debunk the conventional wisdom
that logically centralized computations do not scale
in terms of domain-level end-to-end Internet routes.

\section{Conclusion}
\label{sec:con}

We proposed using IXPs for stitching inter-domain paths under the control 
of centralized routing brokers, which provide paths with end-to-end guarantees
for mission-critical applications. 
We considered a novel abstraction of the Internet topology: the IXP multigraph.
Based on our study using extensive peering datasets, we evaluated the potential of 
IXP-based pathlet stitching in the following ways.
\first In terms of IP address coverage, we showed that even a small deployment ($\sim$5 IXP anchors)
could directly cover a high fraction of the Internet IPv4 address space.
\second In terms of AS-level path diversity, we showed the potential of generalized
routing policies applied on the dense IXP multigraph. We observed an increase of at least one order
of magnitude in path diversity, \ie multiplicity of edge-disjoint paths,
as compared to BGP inter-domain routing practices.
\third We exhibited the importance of having suitable path sampling algorithms that take advantage
of the richness of the multigraph. We further evaluated the
performance and applicability of diverse algorithmic variants---online, offline and hybrid---for
different traffic requirements and graph scales; we have shown that centralized routing variants work
efficiently on the global multigraph view. Lastly, we placed our analysis within the scope of a demanding
application, namely telesurgery, and highlighted open challenges.

As supported by this multi-faceted evaluation
of the potential of CXPs, we believe that providing guaranteed
inter-domain services is not anymore as intractable
as it has been in the past. The flattening Internet topology 
and the emergence of SDN provide new avenues for
innovation on CXP-like approaches. In our on-going
work we investigate ways to kick-start CXP markets. In particular, 
our goal is to still provide better than best effort paths across the Internet, even
when major IXPs or many ISPs do not participate yet in the market.

\section{Acknowledgments}
This work was partly funded by European
Research Council Grant Agreement no. 338402.

\bibliographystyle{}

\setlength{\itemsep}{0pt}
\bibliography{main}
}

\end{document}